\documentclass[prl,twocolumn]{revtex4}
\usepackage{epsfig,color,amsfonts,amsmath,amsthm,comment,natbib}

\usepackage{graphicx}
\usepackage{physics, xcolor}
\usepackage{tikz}

\newcommand{\ltwid}{\mathrel{\raise.3ex\hbox{$<$\kern-.75em\lower1ex\hbox{$\sim$}}}}
\newcommand{\gtwid}{\mathrel{\raise.3ex\hbox{$>$\kern-.75em\lower1ex\hbox{$\sim$}}}}

\newcommand{\bea}{\begin{eqnarray}}
\newcommand{\eea}{\end{eqnarray}}
\theoremstyle{remark}

\newcommand{\old}{\color{black}}

\begin{document}

\bibliographystyle{apsrev}
\title{Quantum wakes in lattice fermions}

\author{Matthew Wampler\footnote{mbw5kk@virginia.edu}, Peter Schauss,  Eugene B Kolomeisky, and Israel Klich\footnote{ik3j@virginia.edu}
}
\affiliation{Department of Physics, University of Virginia, Charlottesville, VA, USA}


\begin{abstract}
The wake following a vessel in water is a signature interference effect of moving bodies, and, as described by Lord Kelvin, is contained within a constant universal angle. However, wakes may accompany different kinds of moving disturbances in other situations and even in lattice systems. Here, we investigate the effect of moving disturbances on a Fermi lattice gas of ultracold atoms and analyze the novel types of wake patterns that may occur. We show how at half-filling, the wake angles are dominated by the ratio of the hopping energy to the velocity of the disturbance and on the angle of motion relative to the lattice direction. Moreover, we study the difference between wakes left behind a moving particle detector versus that of a moving potential or a moving particle extractor. We show that these scenarios exhibit dramatically different behavior at half-filling, with the "measurement wake" following an idealized detector vanishing, though the motion of the detector does still leaves a trace through a "fluctuation wake." Finally, we discuss the experimental requirements to observe our predictions in ultracold fermionic atoms in optical lattices.
\end{abstract}
\maketitle

\section{Introduction}
Many signature effects of classical hydrodynamics have counterparts in quantum systems and serve to provide intuition as well as a spectacular source for interesting new physical situations. Due to the absence of internal scale in hydrodynamics, it can be applied for physical scenarios of vastly different scales. For example, relativistic hydrodynamics has been successfully used to explain collective effects in heavy-ion collisions at RHIC and LHC \cite{jaiswal2016relativistic}. On the other hand, studies of hydrodynamic-like effects in strongly interacting electron systems show unexpected effects due to their similarity to viscous fluids. For example, ref. \cite{guo2017higher} shows that in certain situations, conductance may exceed Landauer's ballistic limit due to viscous effects, while ref. \cite{avron2006adiabatic} demonstrates that slow "swimming" in a Fermi gas is of a topological nature, and can be fine-tuned to be done without dissipation.

Another interesting example of a hydrodynamics inspired study is the investigation of wake waves produced as a response to a moving potential interacting with a two-dimensional electron gas, recently described in \cite{kolomeisky2018kelvin}. There, it was pointed out that the pattern formed is determined by a Mach number and has similarities to Kelvin wakes in water and to Mach shock waves following a supersonic projectile. This behavior can be traced to the coherent interference between plasma excitations in the medium, with a dispersion which is water-like $(\omega(q)^2\propto q)$ at long wavelengths.  A related effect, Cerenkov radiation due to a moving charge in a dielectric has also been studied extensively, most recently in photonic crystals where a host of new variations on the effect have been uncovered where, for example, the direction of radiated energy can be flipped see e.g. \cite{luo2003cerenkov,chen2011flipping}. 

Here, we consider an altogether different system and, with it, a new set of non-equilibrium problems. We examine the discrete-time steady-state generated by the interaction of different types of disturbances, as described below, with fermions on a lattice, as the disturbances move from site to site. Thus, the discrete time, the lattice and the many-body nature of the system play essential roles in the definition of our model. We find that non-classical disturbances may yield a drastically different response.  The case in point is that of a moving quantum particle detector, which in this context has no classical counterpart. In addition, we study a moving particle extraction site, in which particles can be ejected out of the system. These two types of disturbances are compared with results we obtain for a moving potential. We note that in recent years, there have been many investigations of measurement-induced dynamics in many-body systems  \cite{mazzucchi2016quantum,blattmann2016conditioned,mazzucchi2016quantumAntiferro,Fuji2020,Ivanov2020,Ashida2017}. Measurement induced dynamics has been also observed in experiments \cite{Ritsch2013,Schaefer2014,Sorensen2018}. Moving detectors have been considered before as well, most notably in describing the Unruh effect \cite{Unruh1976}, where a uniformly accelerated detector observes a thermal radiation in vacuum. However, the question we consider here is, to the best of our knowledge, completely new: what type of a steady state density pattern will a moving detector leave behind when measuring particle densities in a Fermi sea. 

The recent progress of quantum simulation with ultracold atoms \cite{Gross2017} makes them an ideal platform for studying these effects.
Here, we focus on cold Fermi gases which became a valuable tool in recent years to study non-equilibrium dynamics in analogy to electronic systems.
Indeed, recently, increasingly sophisticated techniques became accessible leading to the measurement of spin dynamics  \cite{Sommer2011,Koschorreck2013,Trotzky2015} and charge transport \cite{Krinner2014,Anderson2019,Xu2019}. In particular, it also became possible to observe spin charge separation in one-dimensional lattice systems \cite{Vijayan2020} and to study spin- and charge transport in the two-dimensional Fermi-Hubbard model in the regime of low temperatures and strong correlations that challenge current theory calculations \cite{Nichols2019,Brown2019}. These experiments demonstrate that ultracold fermionic atoms are an effective platform for quantum simulation of non-equilibrium phenomena even beyond the capabilities of exact calculations.

\begin{figure}[h!]
	    \centering
\includegraphics[width = 0.5\textwidth]{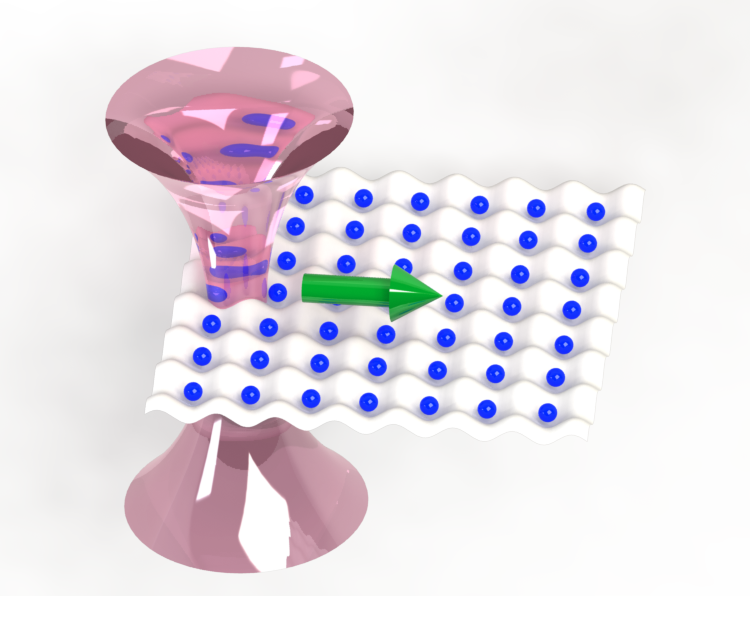}
	    \caption{A lattice of cold atoms interacting with a moving disturbance. The disturbance can be an applied potential, a detector or an extractor. The blue dots represent the fermionic atoms and the red focussed laser beam illustrates the disturbance moving into the direction of the green arrow.} \label{fig:MovingTip}
	\end{figure}

Classically, the universality of wakes following moving ships has been characterized by Kelvin's seminal result, that a (gravity) wake behind a moving ship in water is delimited within a constant angle  $39^{\circ}$, irrespective of the ship's velocity \cite{kelvin1887ship}. Recent results emphasize finite-size effects \cite{darmon2014kelvin} through the dependence on the Froude number $v/\sqrt{g L}$ of a moving pressure source traveling at velocity $v$ of length $L$, where $g$ is the gravitational acceleration constant \cite{colen2019kelvin}. Here, we study wakes  created by point disturbances  moving through a Fermi lattice gas including the quantum effects (Fig. \ref{fig:MovingTip}).  Our results depend on both the velocity and the angle with respect to the Bravais lattice directions, as well as on the type of disturbance.  Concretely, we consider a tip traveling through a lattice of cold fermionic atoms, interacting with an atom on a lattice site, and then during a time $\tau$ moving on to the next site. 

We find several unexpected results. For example, we observe a dramatic difference between the wakes of a moving particle detector and a traveling potential disturbance.  In particular, on the square lattice at half filling the detector wake vanishes identically due to particle hole symmetry at any temperature. Another surprising result is that, at half-filling, the wake formed by a "particle" extractor is independent of temperature. To find an analytic form for the wake left behind a moving
potential we use a co-moving steady-state equation and employ a strategy of identifying nodal lines where the (co-moving)
disturbance is exactly zero, in contrast to most treatments of water wakes, which seek for extrema, i.e.  troughs and crests. Due to the scale inherent in the lattice structure, our wakes depend explicitly on the time $\tau$ characterizing the effective speed of the moving tip, compared to the hopping energy $t_{hop}$ of the fermions in a tight-binding lattice.

\begin{figure}
 \centering
\includegraphics[width = 0.5\textwidth]{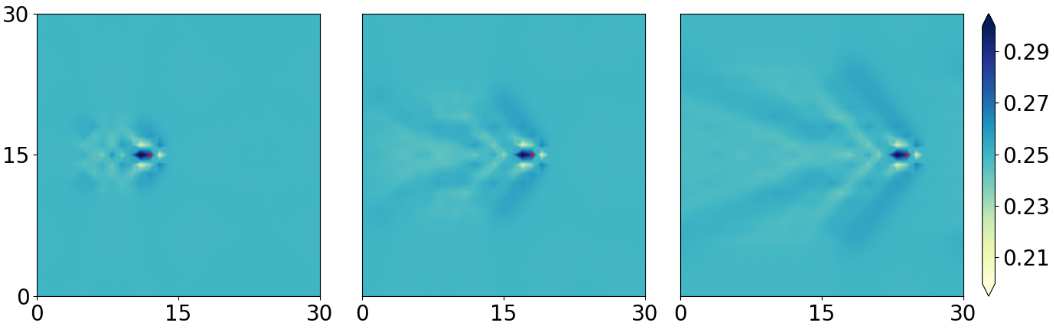}%
\caption{Snapshots of the wake developing following a moving particle detector at quarter filling. Plotted are the local particle densities of each lattice site as given by the color bar on the right. The location of the disturbance is marked by a red dot. The simulations are done by iteration of free evolution (Eq. \eqref{K_Free}) interspersed with interactions with a particle detector (Eq. \eqref{Kdetect}) beginning from the ground state of free fermions on a $30\times 30$ lattice. The detector was initialized at position $(8,15)$ and moves horizontally to the next site during time $1/(3.4~ t_{hop})$, where $t_{hop}$ is the hopping parameter of the free evolution.}\label{filling075Period3detect}
\end{figure}

\begin{figure}
 \centering
\includegraphics[width = 0.5\textwidth]{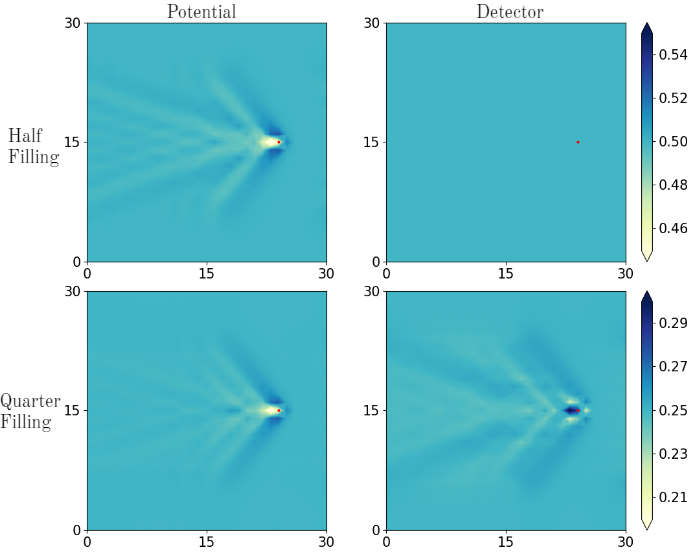}
\caption{Comparison of density plots for wakes of a moving point potential (left) with a moving detector (right).  The detector/point potential was initialized at position $(8,15)$ and moves horizontally to the next site during time $1/(3.4~ t_{hop})$, where $t_{hop}$ is the hopping parameter of the free evolution.  Snapshots are taken after 18 time steps, when the full wake pattern has been sufficiently developed.  At half filling the difference is most dramatic (top) but differences remain at quarter filling (bottom).}\label{DetectVsPotentialvsFilling}
\end{figure}

To describe these effects of dynamics in many-particle quantum systems we use the non-equilibrium framework derived in \cite{klich2019closed}. This framework allows for the study of a variety of non-equilibrium problems including particle detection and injection/extraction. It was shown in \cite{klich2019closed} that in certain statistical mechanics problems, which we detail in the Formalism section, it is possible to make a
systematic connection between the evolution of $n$ body density functions
with $n+1$ density functions, similar in spirit to the Bogoliubov-Born-Green-Kirkwood-Yvon
(BBKGY) hierarchy, which is the essential structure leading to the
Boltzmann equation for single particle densities from higher order
correlation functions (see, e.g. \cite{bonitz1998quantum}).  This approach allows the buildup of tractable non-equilibrium problems utilizing combinations of four elementary operations: detection, particle injection, particle extraction and free evolution.
While some of these ideas have been applied to problems in 1D (e.g.  driven and dissipative XX spin chains \cite{temme2012stochastic} and steady states of a driven hopping model \cite{klich2019closed}), here we study an essential 2D problem: the emergence of wakes behind moving objects interacting with a Fermi sea. In particular, we discuss the difference between the motion of a detector, particle extractor, and a potential in detail. The approach of  \cite{klich2019closed} allows for an efficient numerical calculation of the dynamics in such problems. An example of the development of a wake a moving detector is described in Fig.~\ref{filling075Period3detect}, while a comparison between a moving detector and potential is provided in  Fig.~\ref{DetectVsPotentialvsFilling} at different filling fractions of a Fermi sea in a 2d hopping model. 

The structure of the paper is as follows. 
We start by briefly introducing the formalism of \cite{klich2019closed}. Next we study the effect of a potential hopping from site to site, solving for the characteristic angles of the traveling pattern.  We then continue to study the motion of a detector and the motion of a particle extractor and compare these with our results for the moving potential. Finally, we suggest an experimental setup to directly observe the wake patterns.

\section{Formalism}
First, we provide a formal description of the system depicted in figure \ref{fig:MovingTip}.  
We will denote by $a_{\bf r}$ the annihilation operator for a fermion at lattice site ${\bf r}$.
 To describe the density distribution we will focus on the two point function, defined as:
\begin{eqnarray}\label{defG}
G_{{\bf r}{\bf r'}}(t)=Tr \rho(t)a_{\bf r}^\dag a_{\bf r'} 
\end{eqnarray}
where $\rho(t)$ is the density matrix at time $t$.
The evolution of $G_{{\bf r}{\bf r'}}$ depends on the problem at hand. Due to the discrete nature of the lattice, we find it natural  to consider the time evolution in discrete time steps $\tau$, pertaining to the time disturbance moves from site to site. After a step in the evolution process, $G(t)\rightarrow  {\cal K}(G)\equiv G(t+\tau)$, where ${\cal K}(G)$ is specified for various processes below.

In general, if the system undergoes Hamiltonian evolution during a time $\tau$, we have 
\begin{eqnarray}
    \rho(t+\tau)={\cal U}\rho(t) {\cal U}^\dag
\end{eqnarray}
where ${\cal U}={\cal T}e^{-{i\over \hbar} \int_t^{t+\tau} {\cal H} (s)ds}$ is the many-body evolution. For a general interacting Hamiltonian, $G_{{\bf r}{\bf r'}}(t+\tau)$ is not determined by $G(t)$ alone and would depend on all higher order correlation functions.

For a non-interacting Hamiltonian, however, the evolution of $G$ does not depend on high order correlations. Let us take the Hamiltonian to be  ${\cal H}(t)=\sum_{{\bf r}{\bf r'}} H_{{\bf r}{\bf r'}}(t) a_{\bf r}^\dag a_{\bf r'}$, where $H$ is an $N\times N$ matrix if there are $N$ fermion sites. The evolution of $G$ from time $t$ to time $t+\tau$ is simplified by the fact that for such a Hamiltonian,
\begin{eqnarray}
    {\cal U}^{\dag}a^{\dag}_{\bf q} {\cal U}=U_{\bf q q' }a^\dag_{\bf q'}.
\end{eqnarray}
where $U_{\bf q q'}=[{\cal T}e^{{i\over \hbar} \int_t^{t+\tau} { H^T} (s)ds}]_{\bf q q'}$ is an $N\times N$ single particle evolution operator \footnote{Throughout the rest of the paper we will use units where $\hbar=1$.}. We therefore find for $G$:
\begin{eqnarray}&\label{K_Free}
G_{{\bf r}{\bf r'}}(t+\tau)=Tr {\cal U}\rho(t) {\cal U}^\dag a_{\bf r}^\dag a_{\bf r'} =Tr \rho(t) {\cal U}^\dag a_{\bf r}^\dag a_{\bf r'} {\cal U} \\ \nonumber & 
={[U G(t) U^{\dag}]}_{{\bf r}{\bf r'}}\equiv [{\cal K}_{U}(G)]_{{\bf r}{\bf r'}} ~. 
\end{eqnarray}
In other words, the matrix $G$ undergoes the  evolution $G(t)\rightarrow U G(t) U^{\dag}$, independent of higher correlation functions.
A few other operations that yield a closed equation for $G$ are possible and described in detail in \cite{klich2019closed}. 

We will use two of the aforementioned operations. We start with the elementary particle detection measurement at a site ${\bf r}$. It is described by the following Krauss map of the many body density matrix:
\begin{eqnarray}\label{mapDetection}
    \rho\rightarrow \hat{n}_{\bf r}\rho \hat{n}_{\bf r}+(1-\hat{n}_{\bf r})\rho (1-\hat{n}_{\bf r})
\end{eqnarray}
where $\hat{n}_{\bf r}=a_{\bf r}^\dag a_{\bf r}$ is the number operator associated with site ${\bf r}$. Note that for fermions, $\hat{n}_{\bf r}$ is a projection operator, and the Krauss map \eqref{mapDetection} describes complete decoherence between the number on the site ${\bf r}$ and other sites. Substituting \eqref{mapDetection} in Eq. \eqref{defG}, fermion detection induces the following map on $G$:
\begin{eqnarray}\label{Kdetect}
G\rightarrow {\cal K}_{detect.~r}(G)=P_{{\bf r}}^{\perp}G P_{{\bf r}}^{\perp}+P_{{\bf r}}GP_{{\bf r}}~.
\end{eqnarray}
where $P_{\bf r}=|\mathbf{r} \rangle \langle\mathbf{r}|$ is the (single-particle) projector on site ${\bf r}$ and $P_{{\bf r}}^{\perp}=I-P_{\bf r}$. An additional operation is an extraction event of a particle at site ${\bf r}$. Such an operation can be described by the Krauss map 
\begin{eqnarray}
      \rho\rightarrow \epsilon(2-\epsilon) a_{\bf r}\rho a^{\dag}_{\bf r}+(1-\epsilon\hat{n}_{\bf r})\rho (1-\epsilon\hat{n}_{\bf r}).
\end{eqnarray}
where $0 \leq \epsilon \leq 1$ describes the efficiency of the extraction procedure. Again, substituting this map in the definition \eqref{defG} we obtain
\begin{eqnarray}\label{Kextract}&
G\rightarrow {\cal K}_{extr.~r}(G)= P_{{\bf r}}^{\perp}G P_{{\bf r}}^{\perp}+ \\ \nonumber & (1-\epsilon ) P_{\bf r} G P_{{\bf r}}^{\perp}+(1-\epsilon )P_{{\bf r}}^{\perp} G P_{\bf r}+(1-\epsilon
   )^2P_{\bf r} G P_{\bf r}.
\end{eqnarray}

In this paper, we combine the three types of maps above to represent the dynamics of a disturbance interacting with a lattice as it moves from site to site.
We prepare the system at an initial state, with a two point function $G(t=0)=G_0$. The disturbance will interact with the system at position $\mathbf{r}=0$, and moves to act on an adjacent point $a \mathbf{ w}$, where it acts again after which it moves to $2 a \mathbf{ w}$ and so on, with the disturbance at the $n$th step acting at position $\mathbf{r}=n a \mathbf{ w}$.  We only consider motion at angles where the tip hits the actual lattice sites, i.e. the direction of motion $\mathbf{ w}$ needs to be a lattice vector.

The evolution of the correlation matrix $G$ between successive time steps is described by the  maps given in Eqs \eqref{K_Free}, \eqref{Kdetect}, or \eqref{Kextract}\textemdash as elaborated in detail in each section.  Let the evolution from time $\tau n$ to $\tau (n+1)$ be given by ${\cal K}_n$ (for example, detection at point $\mathbf{r}=n a \mathbf{ w}$ using \eqref{Kdetect}, followed by non-interacting evolution for a time $\tau$ using \eqref{K_Free}). The evolved system at time $n \tau$ will therefore have the correlation matrix:
\begin{eqnarray}
   G(n\tau)= {\cal K}_{n}({\cal K}_{n-1}(....{\cal K}_1(G_0)....)),
\end{eqnarray}
and in particular, the local density change compared to the initial density is 
\begin{eqnarray}&
  \delta G_{{\bf r}{\bf r}}  \equiv Tr (\rho(n\tau)a_{\bf r}^\dag a_{\bf r})-Tr (\rho(0)a_{\bf r}^\dag a_{\bf r})=\\ \nonumber & G_{{\bf r}{\bf r}}(n\tau)-G_{{\bf r}{\bf r}}(0)
\end{eqnarray}
we follow these dynamics numerically, explicitly affecting the iteration procedure for each of the cases, as explained below, and compare the results to the co-moving steady state which we now define. \old

For a moving disturbance, a steady state can only be formed in the co-moving frame. 
Consider an elementary operation $G\rightarrow {\cal K}(G)$ (for example detection, followed by non-interacting evolution for a time $\tau$), which is then repeated, but shifted in space by the vector $a{\bf w}$. Let $S$ be the translation operator along the direction of motion $\mathbf{w}$, via $S^{\dag}~\mathbf{r}=\mathbf{r}+a \mathbf{w} $.
We then define a steady state for the correlation matrix in the co-moving frame via the requirement that
\begin{eqnarray}\label{steadystateGeneral}
G_{steady}=S^{\dag}{\cal K}(G_{steady})S,
\end{eqnarray}
namely $G_{steady}$ is invariant under the combination of the operation ${\cal K}$ and moving to the next site. To identify relevant steady states, we will seek solutions to \eqref{steadystateGeneral} in the vicinity of  states associated with an unperturbed system. Indeed, as we shall see, the nature of the steady states depends both on the form of the dynamics ${\cal K}$ as well as on the initial background state (for example, a Fermi gas at different filling fractions). 

We note that, for simplicity, we focus here on initial states suitable for non-interacting systems.  We emphasize that the formalism is valid also for systems prepared in an
interacting state as long as the subsequent dynamics is well
approximated by non-interacting evolution.
\old


\section{A Moving Potential}
    Consider a tip traveling along the lattice, in a direction $\mathbf{w}$ taking a time $\tau$ to move between two sites. We approximate this process as a discrete process, where a potential $V$ hops from site to site, remaining a time unit $\tau$ at each site.
    For the purposes of this paper, we will focus on the simplest case of a square lattice with nearest neighbor hopping as the free evolution, i.e. ${\cal H}_0 = - t_{hop} \sum_{\langle{\bf r},{\bf r'}\rangle}a_{\bf r}^\dag a_{\bf r'}$ with single particle energies $\varepsilon(k) = -2 t_{hop} \left[ \cos{(k_x a)} + \cos{(k_y a)} \right]$ where $a$ is the lattice spacing. We will take the tip potential at a fixed reference point $\mathbf{r}_0$ to be ${\cal V}=V a_\mathbf{r_0}^\dagger a_\mathbf{r_0}$. We will mostly concentrate on half filling in this section.
    
\begin{figure}
    \centering
    \includegraphics[width = \linewidth]{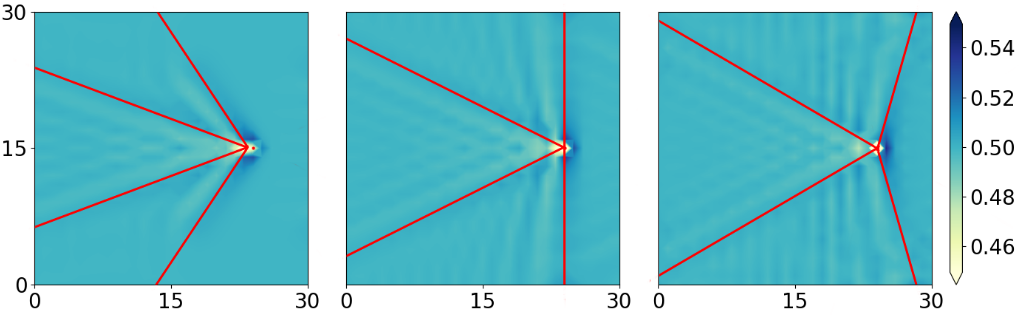}
    \caption{Density plots for varying velocities of a moving potential. From left to right, velocities $\alpha = 1.7$, $1.0$, and $0.7$. Red lines represent the angles given by Eq.~\eqref{Eq. Pot analytic wake geometry gen angle}. Each line corresponds to the solution for a given quadrant in Fig. \ref{fig:Fermi-surface at half-filling}.  Note the forward pointing cone is a result of a forward d-wave radiation when the source is moving slowly.  }
    \label{PotentialWakes1}
\end{figure}

\begin{figure}
    \centering
    \includegraphics[width = \linewidth]{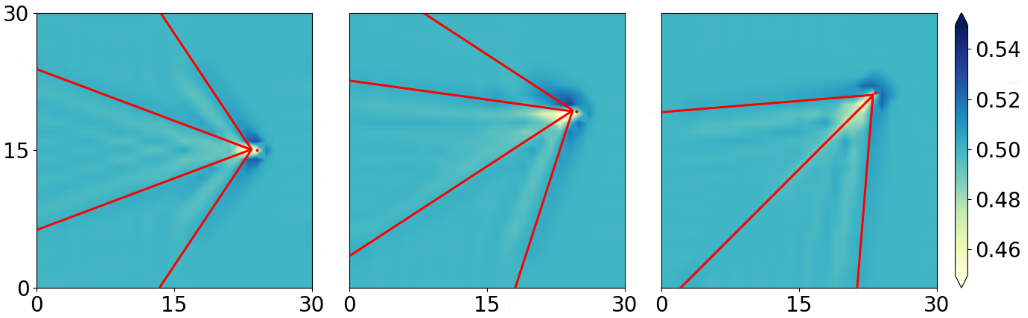}
    \caption{Density plots for varying the angle of a moving potential compared to the lattice vectors. A potential moving at 0, 23, and 45 degrees with respect to the lattice. Here, $\alpha = 1.7$.  A smeared potential (Gaussian with half a lattice spacing width) is used instead of a point potential to include effects when the tip is not precisely on a lattice site.  It is argued in the appendix that the wake geometry is unaffected by this change away from a point potential. Red lines represent the angles given by Eq.~\eqref{Eq. Pot analytic wake geometry gen angle}.  Note that for motion at 45 degrees the angles for quadrants 1 and 4 in Eq.~\eqref{Eq. Pot analytic wake geometry gen angle} coincide  reducing the number of lines to 3.}
    \label{fig:Angles}
\end{figure}

  We describe below the wake formed behind a point potential moving at a general speed and angle with respect to the lattice.  We begin by summarizing the main results of this section before providing full derivations. Fig. \ref{PotentialWakes1} shows the simulation of the wake pattern formed by evolving the system in real time following a successive application of the tip along a horizontal line moving at various speeds. Fig. \ref{fig:Angles} represents the simulation of the wake formed by similarly evolving the system in real time except with the tip moving at several different angles with respect to the lattice.   Denoting 
  
  \begin{equation}
  \alpha = \frac{1}{2 \tau t_{hop}}
  \label{eq: alpha definition}
  \end{equation}
  
  we use Eq. \eqref{steadystateGeneral} to find that the angles of lines of zero disturbance are described by 
    \begin{eqnarray}\label{mainAngles}
  \frac{r_y}{r_x} = \frac{1+ w_y \alpha}{\pm 1+ w_x \alpha}~~;~~  \frac{r_y}{r_x} =\frac{-1+ w_y \alpha}{\pm 1+ w_x \alpha}
    \end{eqnarray}
These "zero disturbance" lines are represented as red lines in the figures, and delineate the shape of the wake openings. As expected, since we are not in a Kelvin regime, the angle depends on the speed of the disturbance and is discussed below. Note, in contrast to the classic Kelvin wakes and potential wakes in a two-dimensional electron gas \cite{kolomeisky2018kelvin}, here the lattice breaks rotational symmetry and the wake pattern changes as the potential path rotates with respect to the lattice.

Before moving on to the derivation, let us comment briefly on the limiting behavior of Eq.~\eqref{mainAngles}.  Note, that as $\alpha \rightarrow 0$, we find $\frac{r_y}{r_x} \rightarrow \pm 1 $, i.e. the two main diagonal directions.  This result is consistent with the expectation that as the velocity vanishes, the moving potential is almost static and will radiate via the underlying D-wave symmetry of the lattice.  
	
	As $\alpha \rightarrow \infty$, i.e. the limit of an extremely fast moving tip, we find that $\frac{r_y}{r_x} \rightarrow \frac{w_y}{w_x}$, in other words, the wake converges onto a line following the disturbance, as any disturbance would not have time to disperse.  Hence, Eq.~\eqref{mainAngles} implies that the  wake will essentially vanish for a potential moving at $\alpha \rightarrow \infty$.  

    The co-moving steady state to be found for our system is described by, Eq.~\eqref{steadystateGeneral}, where ${\cal K}(G)=e^{i \tau (H_0 + V)} G e^{-i \tau (H_0 + V)}$ where $H_0=-t_{hop}\sum_{\langle {\bf r},{\bf r}'\rangle}\ket{{\bf r}}\bra{{\bf r}'} $ is the unperturbed single particle Hamiltonian and $V=V\ket{\mathbf{r_0}}\bra{\mathbf{r_0}}$ is the tip potential at some initial reference point $\mathbf{r}_0$. Namely:
	\begin{gather}\label{NSS_G_U}
S^\dagger e^{i \tau (H_0 + V)} G e^{-i \tau (H_0 + V)} S  = {G} 
	\end{gather}
	
	In general, equation \eqref{NSS_G_U} admits infinitely many solutions for $G$. In particular, any correlation matrix $G$ that satisfies: 
\begin{eqnarray}
[G,S^\dagger e^{i \tau (H_0 + V)}]=0
\end{eqnarray}
will automatically be a co-moving non-equilibrium steady state. In the physical scenario we are interested in, however, we have an initial reference state, the correlation matrix $G_0$ of the unperturbed Fermi system, and in the following we consider the wake as a weak perturbation on this state, allowing us to analytically establish the dominant behavior of the wake pattern. 
	
Since we are only perturbing the free evolution by a small potential, we can assume the steady state $G$ will be close to the steady state of free evolution, $G_0$, where $
G_{0{\bf r}{\bf r'}}=\langle a_{\bf r}^\dag a_{\bf r'} \rangle_{equilibrium}$.  Thus, we write $G=G_0 +\delta G$ where $\delta G$ is assumed a small perturbation. Since $H_0$ is translation-invariant, we write the co-moving non-equilibrium steady state equation in momentum space as:
	\begin{eqnarray*}&
	e^{i a \mathbf{w} \cdot (\mathbf{k} - \mathbf{k'})} \mel{\mathbf{k}}{e^{i \tau (H_0 + V)} (G_0 + \delta G) e^{-i \tau (H_0 + V)}}{\mathbf{k'}} = \\ \nonumber & \mel{\mathbf{k}}{G_0 + \delta G}{\mathbf{k'}}
	\end{eqnarray*}
Substituting lowest order perturbation theory, keeping terms up to linear order in $V$ and $\delta G$, we find that the real space density disturbance, at zero temperature is given by:
	\begin{eqnarray}&\label{eq:deltaGfirst}
		\mel{{\bf r}}{\delta G}{{\bf r}} = 
		\frac{V \tau a^4}{(2 \pi)^4}\iint_{-{\pi\over a}}^{\pi\over a} d\mathbf{k} d\mathbf{k'}
	A(\mathbf{k},\mathbf{k'})R(\mathbf{k},\mathbf{k'},\mathbf{w}) \label{eqn:Pot Integral Before Constraints} \\ & \nonumber
		e^{i ({\bf r}_0-{\bf r}) \cdot (\mathbf{k} - \mathbf{k'})}
	 \left[ \Theta(\varepsilon_f - \varepsilon(\mathbf{k})) - \Theta(\varepsilon_f - \varepsilon(\mathbf{k'})) \right] 
	\end{eqnarray}
	where:
		\begin{eqnarray}&
	A(\mathbf{k},\mathbf{k'})=
		{  \frac{e^{i \tau \left[ \varepsilon(\mathbf{k}) - \varepsilon(\mathbf{k'}) \right]} - 1}{\tau(\varepsilon(\mathbf{k}) - \varepsilon(\mathbf{k'}))}  }, 	\label{eqn:defA} \\ &  
		R(\mathbf{k},\mathbf{k'},\mathbf{w})	={ \frac{1}{1 - e^{-i a \mathbf{w} \cdot (\mathbf{k} - \mathbf{k'})} e^{-i \tau \left[ \varepsilon(\mathbf{k}) - \varepsilon(\mathbf{k'}) \right]}}  } 
		\label{eqn:defR},
	\end{eqnarray}
and $\varepsilon_f$ is the Fermi energy.
Derivation details can be found in the appendix.
	
Our main objective now is to compute the large-scale features of the resulting pattern, namely the typical angle that appears in the wake pattern. As in the case of the original Kelvin wake, which is typically derived by a stationary phase method, the present treatment requires careful consideration of the dominant contribution to the density variation [Eq.~\eqref{eqn:Pot Integral Before Constraints}]. The terms $A$ and $R$ in Eq.~\eqref{eqn:Pot Integral Before Constraints} will provide us with regions that are particularly important for the integral over $\mathbf{k}$ and $\mathbf{k'}$. Due to the Fermi functions, we can write Eq.~\eqref{eq:deltaGfirst} as: 
 	\begin{eqnarray}		\label{eqn: after constraint 3} &
	    \mel{\mathbf{r}}{\delta G}{\mathbf{r}} =
		\frac{2 V \tau a^4}{(2 \pi )^4}\int_{\varepsilon(\mathbf{k})>\varepsilon_f} \int_{\varepsilon(\mathbf{k'})<\varepsilon_f} d\mathbf{k} d\mathbf{k'} \\ \nonumber &
		\Re{	A(\mathbf{k},\mathbf{k'})R(\mathbf{k},\mathbf{k'},\mathbf{w})e^{i ({\bf r}_0-{\bf r}) \cdot (\mathbf{k} - \mathbf{k'})}}
	\end{eqnarray}
Note that $|A(\mathbf{k},\mathbf{k'})|<1$ ({this follows from the inequality $|e^{i\theta}-1|\leq |\theta|$ }) and is dominated by $\mathbf{k}$,$\mathbf{k'}$ near $\varepsilon(\mathbf{k}) - \varepsilon(\mathbf{k'}) = 0$. We thus see that in contrast to the measurement and extraction wakes considered next, the integral is dominated by momenta near the Fermi surface since we can take such momenta to satisfy both conditions $\tau(\varepsilon(\mathbf{k}) - \varepsilon(\mathbf{k'})) \ll 1$ and $\varepsilon(\mathbf{k}) > \varepsilon_f$ and $\varepsilon(\mathbf{k'}) < \varepsilon_f$.

\begin{figure}[h!]
	    \centering
	    \includegraphics{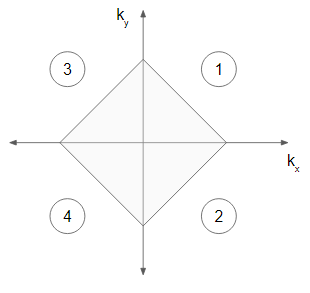}
	    \caption{Fermi surface at half-filling. At half filling all states with momenta $k_x$ and $k_y$ within the diamond shape are occupied. For the calculation we consider the four quadrants separately.} \label{fig:Fermi-surface at half-filling}
	\end{figure}

We will henceforth consider the situation at half filling. Looking at the Fermi surface for our system, we break up the expansion around the Fermi surface into four quadrants given in Fig.~\ref{fig:Fermi-surface at half-filling}.  Close to the Fermi lines,  we will use the variables $\delta_y$ and $\delta_{y'}$ instead of $k_y,k'_y$ as the small shifts away from the Fermi surface. Explicitly,
	\begin{equation}
	    \begin{tabular}{c | c | c}
	       Quadrant & ~ & ~ \\
	       \hline
	         $1$ & $k_y = \frac{\pi}{a} - k_x + \delta_y$ & $ {k'}_y = \frac{\pi}{a} - k'_x + \delta_{y'}$\\
	         $2$ & $k_y = -\frac{\pi}{a} + k_x + \delta_y$ & ${k'}_y = -\frac{\pi}{a} + k'_x + \delta_{y'}$\\
	         $3$ & $k_y = \frac{\pi}{a} + k_x + \delta_y$ &  ${k'}_y = \frac{\pi}{a} + k'_x + \delta_{y'}$\\
	         $4$ & $k_y = -\frac{\pi}{a} - k_x + \delta_y$ &  ${k'}_y = -\frac{\pi}{a} - {k'}_x + \delta_{y'}$\\
	    \end{tabular}
	    \label{eqn: k lines on fermi surface}
	\end{equation}
	
Let us now concentrate on $R$ in Eq. \eqref{eqn:Pot Integral Before Constraints}. This term diverges when 
	\begin{eqnarray}
	\tau [\varepsilon(\mathbf{k}) - \varepsilon(\mathbf{k'})] + a \mathbf{w} \cdot (\mathbf{k} - \mathbf{k'}) = 2\pi n 
	\end{eqnarray} 
for $n$ integer. Here we concentrate on the $n=0$ contribution which already recovers some basic features of the wake pattern, and leave the analysis of $n\neq0$ contributions for a future work. The equation can also be interpreted as a Mach-Cherenkov-Landau condition \cite{Carusotto2013} for the momenta emitted by the  wake due to creating a particle-hole excitation of momentum $\mathbf{K}=\mathbf{k} - \mathbf{k'}$. Perhaps a more familiar way to write the condition is:
	\begin{eqnarray}
\Omega(\mathbf{K})+\mathbf{K}\cdot \mathbf{V} =0
	\end{eqnarray} 
where $V=\tau^{-1}\alpha a \mathbf{w}$, and $\Omega(K)=\nabla_{\mathbf{k}} \epsilon(\mathbf{k})|_{k_F}\cdot \mathbf{K}$.

For the square lattice, we have 
	\begin{eqnarray}
	    \label{eqn: energy is change in k} &
	    \alpha a \mathbf{w} \cdot (\mathbf{k} - \mathbf{k'}) = \\ \nonumber & \cos{(k_x a)} + \cos{(k_y a)} - \cos{(k'_x a)} - \cos{(k'_y a)}
	\end{eqnarray}
	where $\alpha$ is defined by Eq. \eqref{eq: alpha definition}.  
	Now, combining the restriction $\tau(\varepsilon(\mathbf{k}) - \varepsilon(\mathbf{k'})) \ll 1$ with Eq.~\eqref{eqn: energy is change in k}, we find that 
	\begin{equation}    \label{Eq. kx k'x relation}
	    k'_x = k_x + w(k_y - k'_y) + \delta_x
	\end{equation}
	where $\delta_x$ is given by $\delta_x \equiv \frac{\tau(\varepsilon(\mathbf{k}) - \varepsilon(\mathbf{k'}))}{a w_x}$ and where $w \equiv \frac{w_y}{w_x}$.  Comparing to Eq.~ \eqref{eqn: k lines on fermi surface} we arrive at:
	\begin{gather}	    \label{eqn: kx shift of k'x}
	    k_x = k'_x + \delta_{x'}
	\end{gather} 
	where $\delta_{x'} \equiv \frac{w (\delta_y - \delta_{y'}) + \delta_x}{1 +  (-1)^b  w}$. Here, $b = 1$ for quadrant 1 and 4 (Fig.~\ref{fig:Fermi-surface at half-filling}).  Otherwise, $b=0$. Note that our treatment of $\delta_{x'}$ as small breaks down when $w$ is close to $1$.  Indeed, when $w_x=w_y$, the constraints on energy together with Eq. \eqref{eqn: energy is change in k} are insufficient to force $k$ and $k'$ to be close, since $(k-k')$ can be large, with both $k, k'$ close to the Fermi surface, and $(k-k')$ perpendicular to the $\mathbf{w}$ vector. While more refined analysis is needed to describe the special point  $w_x=w_y$ exactly, here we simply observe numerically that our treatment works well for $w_x<w_y$ and $w_y>w_x$, and that the wake pattern change is continuous at $w_x=w_y$ and is well described by our Eq. \eqref{mainAngles}.
	
	
We now combine Eqs.~\eqref{eqn: kx shift of k'x}, \eqref{eqn: k lines on fermi surface}, and 
\eqref{eqn: energy is change in k}, and expand in small $\delta_{x'},\delta_y,\delta_{y'}$ to second order. Solving those, we can relate $\delta_{x'}$ and $\delta_{y'}$ to $\delta_y$ and solve for $ k_x - k'_x$ and $k_y - k'_y$ as
    \begin{eqnarray}&
        k_x - k'_x =\\ \nonumber & \frac{2 \left(w_x \alpha + \sin{(a k_x)} \right) (w_y \alpha + (-1)^{b+1} \sin{a k_x})}{(w_x+ (-1)^b w_y) \alpha \left[(w_x+ (-1)^{b+1} w_y) \alpha + 2 \sin{(a k_x)} \right]}   \end{eqnarray}
        and
            \begin{eqnarray}
            &
        k_y - k'_y = \\ \nonumber &
        \frac{2 \left(w_x \alpha + \sin{(a k_x)} \right)^2}{(w_x+ (-1)^b w_y) \alpha \left[(w_x+ (-1)^{b+1} w_y) \alpha + 2 \sin{(a k_x)} \right]}
    \end{eqnarray}
 
If we assume our potential is at site ${\bf r}_0=(0,0)$, the term $e^{i ({\bf r}_0-{\bf r}) \cdot (\mathbf{k} - \mathbf{k'})}$ in Eq. \eqref{eqn: after constraint 3} becomes
	\begin{gather}
	    e^{i ({\bf r}_0-{\bf r}) \cdot (\mathbf{k} - \mathbf{k'})} \rightarrow e^{-i \mathbf{r} \cdot (\mathbf{k} - \mathbf{k'})} = e^{-i \delta_y B}	\end{gather}
	    where 
	\begin{eqnarray}\label{Bdef}
    & B = \frac{2 \left[w_x \alpha + \sin{(a k_x)} \right]}{(w_x+ (-1)^b w_y) \alpha \left[(w_x+ (-1)^{b+1} w_y) \alpha + 2 \sin{(a k_x)} \right]} \\ \nonumber & \times \left[ r_x (w_y \alpha + (-1)^{b+1} \sin{a k_x}) + r_y (w_x \alpha + \sin{a k_x}) \right]
    \label{eq. A definition}
\end{eqnarray}
	 greatly reducing our momentum space integral to two coordinates, $k_x$ and $\delta_y$. Note that when the denominators in $A,R$  in Eq.~\eqref{eqn: after constraint 3} vanish, the leading behavior of the combination $A R$ is real, we arrive at:
	\begin{eqnarray}\label{EqIntegralApprox}
	    \mel{\mathbf{r}}{\delta G}{\mathbf{r}} \propto
		-\frac{2 V \tau a^4}{(2 \pi)^4} \sum_{\cal Q} \Re{\iint_{\cal Q} dk_x d\delta_y e^{i \delta_y B}}~,
		\label{Eq. Pot Real Space All Quadrants}
	\end{eqnarray}
	where $\cal Q$ is the set of four quadrants in Fig.~\ref{fig:Fermi-surface at half-filling}. We have checked numerically that the integral \eqref{EqIntegralApprox} indeed captures the main wake pattern of the moving potential well. Our next task is to use Eq.~\eqref{EqIntegralApprox} to find the main wake angles.
	
	
	We now estimate analytically the main angles involved in the wake pattern left behind the moving potential.  In the case of water wakes, we are interested in the wavefronts, which are lines of maximal disturbance. Here, we find that a more direct approach is to look instead for lines of zero disturbance, i.e. $\mel{\mathbf{r}}{\delta G}{\mathbf{r}} = 0$.  We will begin by looking at the effects of individual quadrants in Eq.~\eqref{Eq. Pot Real Space All Quadrants}.  Integrating over $\delta_y$ and looking first at quadrant~$1$, we find
	\begin{eqnarray}	\label{Eq. Pot Real Quadrant 1} 
	    \mel{\mathbf{r}}{\delta G}{\mathbf{r}}  &\propto
		-\frac{2 V \tau a^2}{(2 \pi)^4} \Re{\int_0^{\pi/a} \int_0^{k_x} dk_x d\delta_y e^{i \delta_y B}}  
		 \\ \nonumber & = \frac{2 V \tau a^2}{(2 \pi)^4} \Im{\int_0^{\pi/a} dk_x \frac{e^{i k_x B} - 1}{B}}
	\end{eqnarray}
 To find the characteristic wake lines, we now look for directions $\mathbf{r}$ such that Eq.~\eqref{Eq. Pot Real Quadrant 1} vanishes.  Assuming that we could treat the equation by a stationary phase method, a condition for Eq.~\eqref{Eq. Pot Real Quadrant 1} vanishing would be that there exists a $k_0$ such that $k_x B \approx (k_x - k_0)^2$. In this case, using the stationary phase approximation around $k_0$ makes the dominant contribution to the integral in Eq.~\eqref{Eq. Pot Real Quadrant 1} real, and $\mel{\mathbf{r}}{\delta G}{\mathbf{r}}$ vanishes.  Specifically, for this to happen, we need $B=0$ and $\frac{d}{d k_x}B = 0$ when evaluated at $k_0$.  Looking at $r_x , r_y \gg 1$, i.e. far away from the potential, the dominating behavior of $B$ (Eq. \eqref{Bdef}) comes from $ r_x (w_y \alpha + (-1)^{b+1} \sin{a k_x}) + r_y (w_x \alpha + \sin{a k_x})$.  Hence, we find the equations:
	\begin{eqnarray} \label{eq: k0 condition 1}
	     r_x (w_y \alpha + (-1)^{b+1} \sin{a k_x}) + r_y (w_x \alpha + \sin{a k_x}) = 0,
	   	\end{eqnarray}  
	     and
		      	\begin{eqnarray} \label{eq: k0 condition 2}
(-1)^{b+1} r_x \cos{a k_x}  +r_y \cos{a k_x} = 0.
	\end{eqnarray}
	Therefore, $\cos{a k_x} = 0$ implying $k_0 = \frac{\pi}{2 a}$. Plugging this into Eq.~\eqref{eq: k0 condition 1} yields 
	\begin{equation}
	   \frac{r_y}{r_x} = \frac{1+w_y \alpha}{1+w_x \alpha}.
	\end{equation}
	Repeating this calculation for the other three quadrants, we find
	\begin{equation}
    \begin{tabular}{c | c}
       Quadrant & Line of $\mel{\mathbf{r}}{\delta G}{\mathbf{r}} = 0$ \\
       \hline
       1 & $\frac{r_y}{r_x} = \frac{1+w_y \alpha}{1+w_x \alpha}$ \\
       2 & $\frac{r_y}{r_x} = \frac{-1+w_y \alpha}{1+w_x \alpha}$ \\
       3 & $\frac{r_y}{r_x} = \frac{1+w_y \alpha}{-1+w_x \alpha}$ \\
       4 & $\frac{r_y}{r_x} = \frac{-1+w_y \alpha}{-1+w_x \alpha}$ \\
    \end{tabular}
    \label{Eq. Pot analytic wake geometry gen angle}
    \end{equation}
    and hence our main result Eq.~\eqref{mainAngles}.
Figures \ref{PotentialWakes1} and \ref{fig:Angles} show agreement between the simulations of the potential wakes and our  Eq. \eqref{Eq. Pot analytic wake geometry gen angle}. 
While the above treatment was perturbative in $V$, our analytical treatment for the angles should hold asymptotically at large distances from the source.  This is because, in that regime, the response to the disturbance is weak regardless of the strength of the perturbation.
Close to the source, the density profile will, in general, not be linear in the strength of the perturbing potential. In particular, at half filling, the requirement $\delta G\ll G_0$ implies that $	\mel{{\bf r}}{\delta G}{{\bf r}}\ll {1\over 2}$. To see the range of validity of the description close to the source, let us consider $\mel{{\bf r}}{\delta G}{{\bf r}}$	as expressed in the integral \eqref{Eq. Pot Real Quadrant 1}. We note that $B$ grows linearly in $|\mathbf{r}|$, and therefore, asymptotically $\mel{\mathbf{r}}{\delta G}{\mathbf{r}}$ oscillates and decays at least as fast as  $|\mathbf{r}|^{-1}$ (consistent with the numerical observations).
The expression \eqref{Eq. Pot Real Quadrant 1} shows that the density profile will be perturbative when  ${V\tau a
\over |\mathbf{r}|}\ll 1$, or, in other words at distances ${|\mathbf{r}|\over a}\gg V\tau $.  
\old

    \section{Moving particle extractor and moving detectors}
We proceed to consider a moving detector or particle extraction from the system. Note, that these processes are non-unitary. In this section we establish the co-moving steady-state for this problem. In particular, we show that in marked contrast with a moving potential, a moving detector at half filling does not generate a wake because of particle-hole symmetry.

We assume that the detection or extraction process is dominant when the tip is at a given site, but quickly weakens as the tip moves away from that site. It is therefore natural to discretize the process in such a way that we have a disturbance at a given site, followed by a free evolution of the system during a time $\tau$ that the tip is traveling to the next site on it's trajectory.
The appropriate transformation rules ${\cal K}(G)$ for detection and extraction are given in Eq.~\eqref{Kdetect} and \eqref{Kextract} respectively.  If we allow for pure detection to happen with probability $p$ (associated with the efficiency of the detector) and similarly extraction protocol with probability $q$, we can combine them, together with the free evolution $U=e^{-i\tau H_0}$ into the general form
\begin{eqnarray}&\label{extraction detection}
G\rightarrow {\cal K}(G)=\\ \nonumber & {\cal K}_{U}((1-p-q)G+p {\cal K}_{detect}(G)+ q {\cal K}_{extract}(G)) )
\end{eqnarray}
which can be written as:
\begin{eqnarray}
G\rightarrow {\cal K}(G)=U^\dagger \left[ G - \gamma \{ G,P \} + \xi  P G P \right] U
\end{eqnarray}
where $\xi=2p+\epsilon^2 q$, $\gamma=p+\epsilon q$, $\{G,P\} \equiv GP + PG$ indicates the anti-commutator, and $P$ is the projection onto a site ${\bf r}_0$ where the tip acts. In particular, pure detection will be described by $q=0$, hence $\gamma=p$ and $\xi=2p$.

\begin{figure}[t] 
        \centering
        \includegraphics[width =\linewidth]{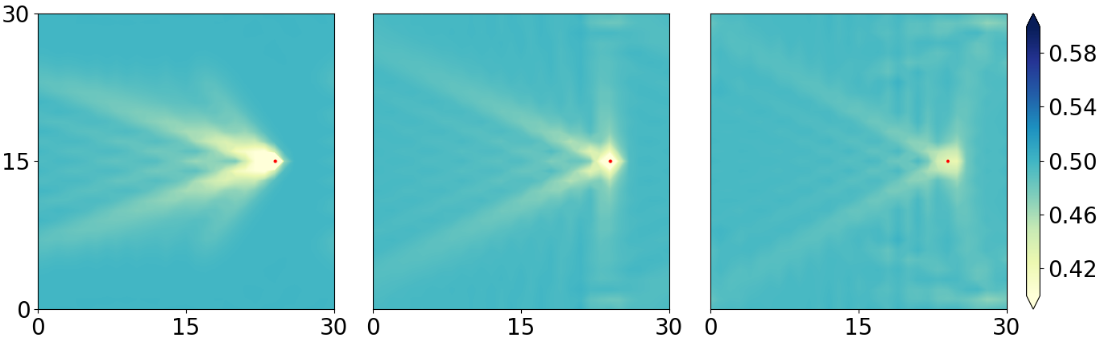}
        \caption{Density plot for varying speeds of a moving particle extractor at half-filling. From left to right, speed $\alpha = 1.7$, $1.0$, and $0.7$ respectively.}
        \label{fig:Removal Speeds}
    \end{figure}

In the next sections we work under the assumption that $p,q\ll 1$ and hence $\gamma\ll 1$. 
The co-moving steady state equation \eqref{steadystateGeneral} now reads:
  \begin{eqnarray}
        \mel{\mathbf{k}}{S^\dagger U^\dagger \left[ G - \gamma \{ G,P \} + \xi  P G P \right] U S}{\mathbf{k'}}\! =\! \mel{\mathbf{k}}{G}{\mathbf{k'}}
        \label{vacuumStart}
    \end{eqnarray}
Written explicitly in momentum space we have:
    \begin{eqnarray*}&
        e^{i a \mathbf{w} \cdot (\mathbf{k} - \mathbf{k'})} e^{i \tau \left[ \varepsilon(\mathbf{k}) - \varepsilon(\mathbf{k'}) \right]} \{ \mel{k}{G}{k'} - \gamma \mel{\mathbf{k}}{ \{ G,P \}}{\mathbf{k'}} + \\ \nonumber &  \xi \mel{\mathbf{k}}{P G P}{\mathbf{k'}} \} = \mel{\mathbf{k}}{G}{\mathbf{k'}}
    \end{eqnarray*}
Assuming that $\gamma \ll 1$, $G \approx G_0 + \delta G$, with $\delta G$ being a small correction, and zero temperature, we find that the local density variation is given by:
    \begin{eqnarray}\label{eq: detection final}
    & \mel{\mathbf{r}}{\delta G}{\mathbf{r}} = \frac{\gamma a^4}{(2\pi)^4} \iint_{-\pi/a}^{\pi/a} d\mathbf{k} d\mathbf{k'}  R(\mathbf{k},\mathbf{k'},\mathbf{w}) e^{i (\mathbf{r_0} - \mathbf{r}) \cdot (\mathbf{k} - \mathbf{k'})} \nonumber
\\ 
      & \left[ \frac{\xi \rho_f}{\gamma} - \Theta(\varepsilon_f - \varepsilon(\mathbf{k})) - \Theta(\varepsilon_f - \varepsilon(\mathbf{k'})) \right]
    \end{eqnarray}
 where $\rho_f$ is the density of fermions in $G_0$ (i.e. the diagonal of the $G_0$ matrix).  \old Note, that like in the potential case, the term $R$ in Eq.~\eqref{eq: detection final} implies that Eq.~\eqref{eqn: energy is change in k} still characterizes a dominant region for the integral. 
    However, Eq.~\eqref{eq: detection final} has two added difficulties when compared with the moving potential case Eq.~\eqref{eq:deltaGfirst}.  First, we no longer have the helpful constraint that $\tau \left[ \varepsilon(\mathbf{k}) - \varepsilon(\mathbf{k'}) \right] \approx 0$.  Secondly, $\varepsilon(\mathbf{k})$ and $\varepsilon(\mathbf{k'})$ can now be on the same side of $\varepsilon_f$ as well as on the opposite side. Nonetheless,  in numerical experiments we have observed that the geometry for a moving potential, Eq. \eqref{Eq. Pot analytic wake geometry gen angle}, does appear to also match with the wake patterns of a moving detector and extractor, as can be observed, e.g., by comparing the wake pattern Fig. \ref{PotentialWakes1} to the extractor pattern Fig. \ref{fig:Removal Speeds} .
    
\begin{figure}[t] 
        \centering
        \includegraphics[width =\linewidth]{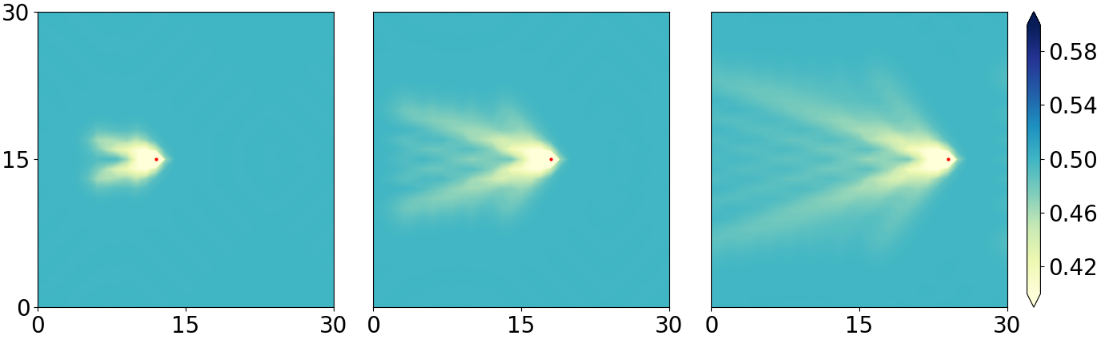}
        \caption{Density plot of a wake developing following a fermion extraction site moving at $\alpha = 1.7$.  The pictures show a steady-state in the comoving frame.}
        \label{fig:Removal Animation}
    \end{figure}

By iterating the evolution equation for the two point function $G$, the wake pattern can be generated numerically. For the case of a particle removal site moving through a half-filled Fermi sea we obtain the images shown in Fig.~\ref{fig:Removal Animation} and \ref{fig:Removal Speeds}. The geometry of the wake patterns is similar to the ones described for the moving potential Fig. \ref{PotentialWakes1}, however the density variation is always negative due to the depleted particles.

    \subsection{Moving detector at half filling}
Particle detection at half filling shows marked contrast with the density wake due to a moving potential. Indeed, due to particle hole symmetry it leaves the average density profile, namely the diagonal of $G$ unchanged. On the other hand, a potential perturbation breaks particle-hole symmetry and generates the density wake described above. 

In fact, we can establish a stronger property, namely:
    \begin{equation}
        \label{eq: particle-hole}
        \mel{\mathbf{r}}{\delta G}{\mathbf{r}}_{\mu} + \mel{\mathbf{r}}{\delta G}{\mathbf{r}}_{-\mu} = 0
    \end{equation}
where $\mel{\mathbf{r}}{\delta G}{\mathbf{r}}_{\mu}$ is obtained by successive applications of ${\cal K}$ from Eq. \eqref{extraction detection} on an initial state $G_0={1\over 1+e^{\beta(H-\mu)}}$, with $q=0$, using an arbitrary choice of measuring site at each step, and in the end subtracting $G_0$. 
In other words, changing the sign of the chemical potential changes the sign of the wake.

\begin{figure}[h]
        \centering
        \includegraphics[width=\linewidth]{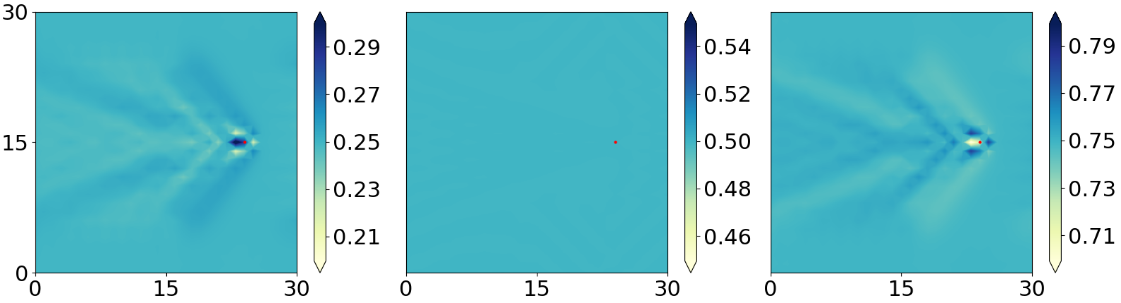}
        \caption{Density plots of a detector moving through a Fermi sea at several filling fractions. From left to right, the detector is moving at $\alpha = 1.7$ for quarter-filling, half-filling, and three-quarter-filling, respectively. The color bar has the same scale but different offset (centered around the initial filling fraction).}
        \label{fig:Moving Detector}
    \end{figure}

An immediate consequence of Eq.~\eqref{eq: particle-hole} is that at the point, where our many-body Hamiltonian has particle-hole symmetry, namely $\mu=0$ i.e. at half filling:
   \begin{equation}
        \label{eq: no det wake half fill}
        \mel{\mathbf{r}}{\delta G}{\mathbf{r}}_{\mu=0}= 0,
    \end{equation}
    i.e. there should be no wake pattern created by a moving detector.  This is shown in Fig.~\ref{fig:Moving Detector} by comparing a detector moving through a half-filled versus a quarter-filled Fermi sea. The image also shows how the quarter filled wake is opposite in sign to the wake generated in the Fermi system at three quarter filling.
    
   A full non-perturbative proof of the remarkable relation Eq.~\eqref{eq: particle-hole} is presented in the appendix. Here for simplicity we  establish Eq.~\eqref{eq: particle-hole} starting from the (zero temperature) perturbative result Eq.~\eqref{eq: detection final} with $\epsilon_F=\mu$ and $\xi=2\gamma$. Note that the sum $ \mel{\mathbf{r}}{\delta G}{\mathbf{r}}_{\mu} + \mel{\mathbf{r}}{\delta G}{\mathbf{r}}_{-\mu} $ is given by  Eq.~\eqref{eq: detection final} with the term in brackets replaced with
    \begin{equation}
        \label{Eq: mu plus minus mu}
        2 - \Theta(\mu - \varepsilon(\mathbf{k})) - \Theta(\mu - \varepsilon(\mathbf{k'})) - \Theta(-\mu - \varepsilon(\mathbf{k})) - \Theta(-\mu - \varepsilon(\mathbf{k'}))
    \end{equation}
    where we used that $\rho_f(\epsilon_F)+\rho_f(-\epsilon_F)=1$. 
    
    Now consider the following map reflecting points about the Fermi surface:
    \begin{eqnarray}
    \mathbf{k} \rightarrow {\cal M} (\mathbf{k}) \equiv (-1^{b},-1^{1+{\cal Q}}) {\pi\over a} - \mathbf{k}
    \end{eqnarray}
    where $\cal Q$ is the quadrant number and $b=0$ if in quadrants 1 or 2 and $b=1$ in quadrants 3,4.
    Note that $\epsilon({\cal M} (\mathbf{k}))=-\epsilon (\mathbf{k})$, and that, 
    $\exp[i \mathbf{r}\cdot({\cal M} (\mathbf{k})-{\cal M} (\mathbf{k}'))]=\exp[-i \mathbf{r}\cdot(\mathbf{k}-\mathbf{k}')]$
 thus, the real part of $R(\mathbf{k},\mathbf{k'},\mathbf{w}) e^{i (\mathbf{r_0} - \mathbf{r}) \cdot (\mathbf{k} - \mathbf{k'})}$ in Eq.~\eqref{eq: detection final} is symmetric under such a transformation. On the other hand, the bracket term Eq. \eqref{Eq: mu plus minus mu} is anti-symmetric under the map ${\cal M}$. The result of $k,k'$ integrations will therefore vanish, establishing Eq. \eqref{eq: particle-hole}.
    
    \old
    
\subsection{A "fluctuation" wake}
\begin{figure}
    \centering
    \includegraphics[width = 0.3333\textwidth]{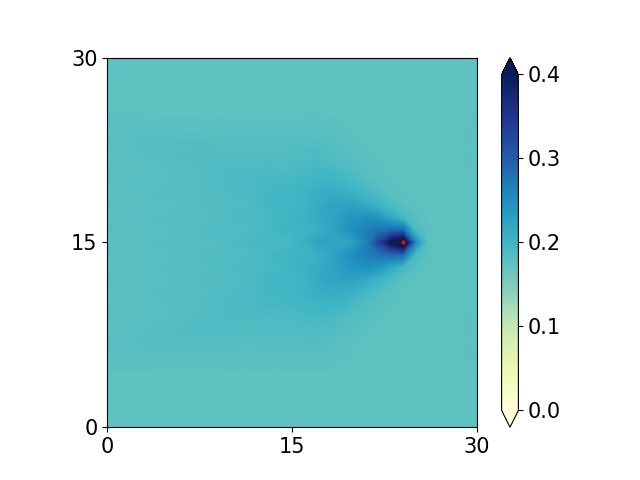}
    \caption{Density of particles in a spatially spread mode centered around each lattice site (given by $n_A({\bf r})$ as defined in Eq. \eqref{Eq: Fluctuations definition}) following a moving detector at half-filling.  Here, $\alpha = 1.7$.}
    \label{fig: Detection fluctuations}
\end{figure}

The above results suggest at first glance that there is no effect of the detector at half filling. In fact, this is not the case!
While the moving detector does not affect the average density at half filling, it does perturb correlations, and thus may be observed through fluctuations. For example, such correlations may be observed by looking at the number of particles in a mode $a_{A({\bf r})}^{\dag}\equiv{1\over \sqrt{A}}\sum_{{\bf r'-r}\in A}a_{\bf r'}^{\dag}$, representing an equal weight superposition in a region lattice neighborhood $A$ of a point ${\bf r}$. We have:
\begin{eqnarray}&
n_A({\bf r})\equiv \langle a_{A({\bf r})}^{\dag}a_{A({\bf r})}\rangle=
{1\over |A|}\sum_{{\bf r'}{\bf r''}\in (A+{\bf r})} G_{{\bf r'} {\bf r''}}= \label{Eq: Fluctuations definition} \\ \nonumber &  {1\over |A|}\sum_{{\bf r''}{\bf r'}\in (A+{\bf r})} {G_0}_{{\bf r'} {\bf r''}}+{1\over |A|}\sum_{{\bf r'}{\bf r''}\in (A+{\bf r})} {\delta G}_{{\bf r'} {\bf r''}}
\end{eqnarray}
We will focus on $A$ being the set of nearest neighbors:
an example of the wake in the density of the $n_A({\bf r})$ is then shown in Fig.~\ref{fig: Detection fluctuations}.
That this wake may be non-zero can be observed by generalizing Eq.~\eqref{eq: detection final} for off-diagonal elements.  In this case, the only change to \eqref{eq: detection final} is that $e^{-i \mathbf{r} \cdot (\mathbf{k} - \mathbf{k'})} \rightarrow e^{-i ( \mathbf{r} \cdot \mathbf{k} - \mathbf{r'} \cdot \mathbf{k'})}$.  Now, combining Eqs.~\eqref{eq: detection final} and Eq.~\eqref{Eq: Fluctuations definition}, we find 
\begin{eqnarray}&
    e^{-i ( \mathbf{r} \cdot \mathbf{k} - \mathbf{r'} \cdot \mathbf{k'})} \rightarrow e^{-i \mathbf{r} \cdot (\mathbf{k} - \mathbf{k'})} \sum_{q,q' \in \kappa} e^{-iqa + iq'a} \label{Eq: Fluctuations exp change} \\ \nonumber & = e^{-i \mathbf{r} \cdot (\mathbf{k} - \mathbf{k'})} (1 - \frac{\varepsilon(\mathbf{k})}{t_{hop}})(1 - \frac{\varepsilon(\mathbf{k'})}{t_{hop}})
\end{eqnarray}

where $\kappa = \{0,k_x,k_y,-k_x,-k_y\}$.

Note, Eq.~\eqref{Eq: Fluctuations exp change} is not symmetric under a reflection of $\varepsilon(\mathbf{k})$,$\varepsilon(\mathbf{k'})$ about the Fermi energy.  This implies that, unlike the diagonal of $G_{r r}$, the wake generated for the modes like $A$ for a detector are non-zero.  

A couple of remarks are in order.\\
(1) While we focused here on the density of the $a_A^{\dag}$ modes as an indicator of correlations, a more natural quantity for an experimental consideration is density-density correlations, and number fluctuations in the region $A$. A preliminary check shows that such number fluctuations will also exhibit a wake, which can be studied by considering the four-fermi correlation generalization of Eq.~\eqref{K_Free} and \eqref{Kdetect}, which give a closed hierarchy of 4 point functions. This calculation will be presented in a future work.
\\
(2) A moving detector at half filling is an interesting example of a "hierarchy" of steady states. 
In this hierarchy, the local average density or "diagonal" of $G$ at half filling is steady for any path a detector makes, and is thus in a steady state. However, the correlations depend on the trajectory of the detector and would, in general, not be in a steady state, moreover the many-body density matrix would not be in a steady state.  It is not hard to construct examples where $G$ is in a steady state, while the many body density matrix is time-dependent.  

    \section{Finite temperature states} 
\begin{figure}[t]
        \centering
        \includegraphics[width = 0.5\textwidth]{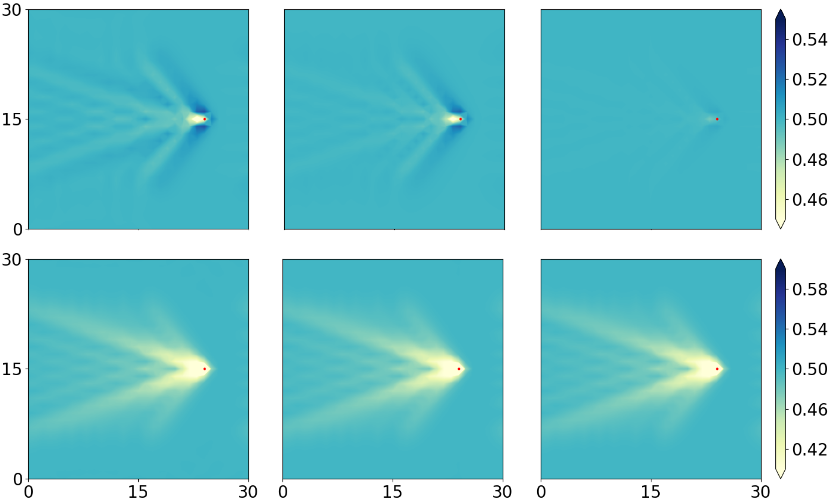}
        \caption{Density plots of potential (top) and extraction (bottom) wakes at half-filling over varying temperature.  From left to right, $\frac{T}{t_{hop}} = 0$, $1$, and $10$.}
        \label{fig:Pot Extr Variable Temp Pictures}
    \end{figure}

\begin{figure}[t]
        \centering
        \includegraphics[width = 0.5\textwidth]{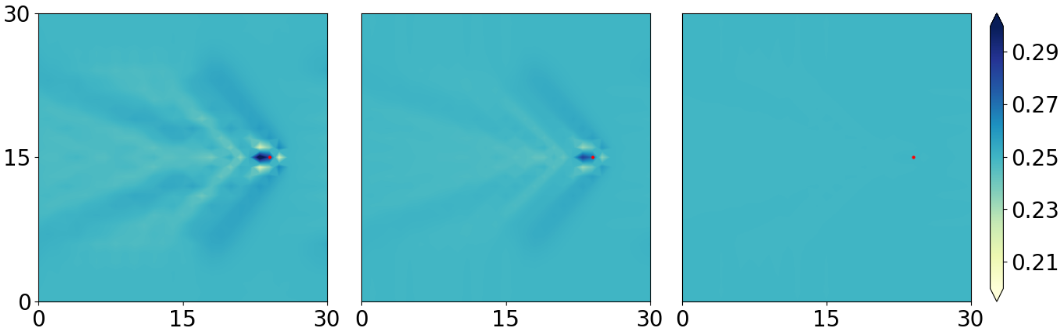}
        \caption{Density plots of detection wakes at quarter-filling and varying initial temperature.  From left to right, $\frac{T}{t_{hop}} = 0$, $1$, and $10$.}
        \label{fig:Det Variable Temp Pictures}
    \end{figure}

    In this section we analyze the effect of a non-zero temperature of the system on our moving disturbances. We assume that the system is prepared initially at finite temperature, and we neglect thermal dissipation on the time scale of the motion of our disturbances. We find that at a generic filling the amplitude of the wakes are decreased, as may be expected on general grounds, i.e. with increased density fluctuations of the background. These results are shown in Figs.~\ref{fig:Pot Extr Variable Temp Pictures} and \ref{fig:Det Variable Temp Pictures}.  Furthermore, we find that, at $\rho_f = \frac{1}{2}$, a moving detector continues to produce no wake at finite temperature.
    Perhaps the most surprising effect we find is that the extractor wake at half filling is temperature independent.  This behavior is striking when compared to the moving potential source, see Fig.~\ref{fig:Pot Extr Variable Temp Pictures}. 
 
At finite temperature, $\mel{\mathbf{k}}{G_0}{\mathbf{k'}} = \delta_{\mathbf{k} \mathbf{k'}} F(\varepsilon(\mathbf{k}))$ instead of $\delta_{\mathbf{k} \mathbf{k'}} \Theta(\varepsilon_f - \varepsilon(\mathbf{k})$, where $F(\varepsilon(\mathbf{k}))$ is the Fermi-Dirac distribution. We find that the finite temperature steady states, are simply obtained by replacing the step functions in equations  \eqref{eq:deltaGfirst} and \eqref{eq: detection final} by Fermi-Dirac functions $F(\varepsilon(\mathbf{k}))$.

Thus, the steady state of $\delta G$ for a moving potential source, Eq.~\eqref{eqn:Pot Integral Before Constraints}, becomes
    \begin{eqnarray}
		& \mel{\mathbf{r}}{\delta G}{\mathbf{r}} = 
		\frac{V \tau a^4}{(2 \pi)^4}\iint_{-\pi/a}^{\pi/a} d\mathbf{k} d\mathbf{k'}
		A(\mathbf{k},\mathbf{k'}) R(\mathbf{k},\mathbf{k'},\mathbf{w}) \nonumber \\ & 
		e^{i (\mathbf{r_0}-\mathbf{r}) \cdot (\mathbf{k} - \mathbf{k'})}
		\left[ F(\varepsilon(\mathbf{k})) - F(\varepsilon(\mathbf{k'})) \right]
		\label{eqn:Pot Integral finite Temp}
	\end{eqnarray}
while for detection/extraction at finite temperature, Eq.~\eqref{eq: detection final} becomes
    \begin{eqnarray}\label{extract_detect_finiteT}
    & \mel{\mathbf{r}}{\delta G}{\mathbf{r}} = \frac{\gamma a^4}{(2\pi)^4} \iint_{-\pi/a}^{\pi/a} d\mathbf{k} d\mathbf{k'}  R(\mathbf{k},\mathbf{k'},\mathbf{w})  e^{i (\mathbf{r_0}-\mathbf{r}) \cdot (\mathbf{k} - \mathbf{k'})} \label{eq: detection final finite temp} \nonumber \\ &  \left[\frac{\xi \rho_f}{\gamma} - F(\varepsilon(\mathbf{k})) - F(\varepsilon(\mathbf{k'})) \right].
    \end{eqnarray}
We can understand the temperature independence of the moving extractor at half filling as follows. Consider  the difference between the moving extractor and moving detector steady state equations (Eq. \eqref{eq: detection final finite temp} with $\xi = 2 \gamma$ and $\xi = \gamma$ respectively), we find:
    \begin{eqnarray}\label{differenceindTemppert}
        &\mel{\mathbf{r}}{\delta G_{det}}{\mathbf{r}} - \mel{\mathbf{r}}{\delta G_{extr}}{\mathbf{r}} = \label{eq: det and siphon difference}\\ & \nonumber
        \frac{\gamma a^4}{(2\pi)^4} \iint_{-\pi/a}^{\pi/a} d\mathbf{k} d\mathbf{k'}  R(\mathbf{k},\mathbf{k'},\mathbf{w}) e^{i (\mathbf{r_0}-\mathbf{r}) \cdot (\mathbf{k} - \mathbf{k'})} \rho_f 
    \end{eqnarray}
    Note, Eq.~\eqref{eq: det and siphon difference} depends only on the density $\rho_f$.  Therefore  the difference Eq.~\eqref{eq: det and siphon difference} is independent of temperature if temperature is varied at a fixed density.
 Since detection creates no wake when $\rho_f = \frac{1}{2}$ at any temperature, this implies that a moving particle extractor is temperature independent at $\rho_f = \frac{1}{2}$.  The result above, Eq. \eqref{differenceindTemppert}, that the difference between the detector and extractor wakes is temperature independent has been done perturbatively in $\delta G$ for illustration purposes. In fact it is possible to establish that
 \begin{gather}
     \frac{d}{dT}(\mel{\mathbf{r}}{G_{det}}{\mathbf{r}} - \mel{\mathbf{r}}{G_{extr}}{\mathbf{r}}) = 0
 \end{gather}
 where $G_{det}$,$G_{extr}$ are the non-perturbative steady states for a moving detector and extractor respectively, as derived in the appendix.  \old
 This result matches simulations of the moving particle extractor as shown in Fig.~\ref{fig:Pot Extr Variable Temp Pictures}. We note, in passing, that numerical checks show that the fluctuation wake is temperature dependent at half filling, even though there is no detector wake.

	
    \section{Discussion of experimental realizations}

Here we consider a setup where the wakes may be explored in experiments with ultracold $^6$Li fermions in optical lattices. Quantum gas microscopes with single-particle and single-site resolution can directly observe the wake structure, as follows. The disturbance can be created by a focused laser beam with a waist on the order of the lattice spacing by employing a high-resolution objective \cite{Weitenberg2011}. Experimental system sizes of more than 30x30 lattice sites have been realized for fermions \cite{Brown2019}.
We have checked by numerical simulations that using a Gaussian smeared potential instead of a point potential does not significantly alter wake geometry. Moreover, the wake pattern is not significantly changed if the initial Gaussian is not centered on a lattice site.

Because of the light mass of $^6$Li the timescales for Hubbard physics are still convenient for lattice spacing of approximately $1\,\mu $m \cite{Vijayan2020} leading to moderate requirements of the objective (NA $> 0.5$).  A dynamically movable disturbance can be implemented, for example, by piezo-actuated mirror mount or an acousto-optical deflector. An experimental run will then start with the preparation of a two-dimensional Fermi-Hubbard system and the movement of the focused laser beam through this system up to a certain position. Finally, the system is frozen by increasing the lattice depth and then imaged via fluorescence imaging. A single realization does not contain enough information to extract the details of the wake pattern. Reaching a density resolution of about 2$\%$ requires averaging about 2500 experimental realizations ($1/\sqrt{n}$) while keeping track of the final position of the disturbance. The required precision and amount of data is comparable to recent experiments at existing quantum gas microscopes \cite{Weitenberg2011,Vijayan2020}. The parameter $\tau$ is proportional to the duration of beam motion, which should be compared to the hopping energy $t_{hop}$. It is possible to swipe the beam at different rates to obtain
values of $\alpha$ spanning the full range from the slow-moving D-wave like wakes to the disappearance of the disturbance at high speeds. We remark that in a finite optical lattice setting, the large scale wake pattern may not have enough time to develop if the speed is such that the co-moving steady state cannot be effectively reached.

We note that all three types of disturbances can be implemented in experiments. A moving potential can be created by a far-detuned laser beam. A moving detector can be realized by a near-resonant laser beam. Scattering of photons at low intensity leads in good approximation to a measurement of the on-site particle density. Last, a particle extractor can be implemented via a defocussed red-detuned optical dipole trap. Caused by the out-of-plane minimum in the potential, atoms will be sucked out-of-plane and will be lost in the experiment.

Beyond cold atoms, we expect the effects we predict to also hold in other systems which can be well described by non-interacting fermions. We note that, for spin-polarized fermions contact-interactions are suppressed at low temperatures and dynamics is dominated by the effects of Pauli exclusion, making our approach particularly effective in this case. Moreover, we emphasize that our treatment is essentially exact, and thus can provide a benchmark for studies of the perturbative effect of interactions.

It is also important to note that while the present discussion is focused on non-interacting systems, the formalism presented in \cite{klich2019closed} is valid also for systems prepared in an interacting state, as long as the subsequent step of unitary evolution while the tip is traveling between sites is well approximated by non-interacting evolution. Thus, a system prepared in a strongly correlated state, such as a Mott insulator, for example, that undergoes a quantum quench where interactions are turned off will still be described by the current method.  
 
 \emph{Acknowledgments.}
The work of I.K. and M.W. was supported in part by the NSF Grant No. DMR-1918207.

    \bibliographystyle{unsrt}

\newpage

\onecolumngrid
\appendix
    

\section{Appendix: The steady state Equations}\label{derivation_steady_states}
(1) {\bf Moving potential.}
Here, we explain the iterative equations for the two point correlation functions and the density in the case of non-interacting evolution. We start with the case of a system interacting with an external potential, no measurements or change in particle number is involved and so the evolution is unitary. In this case, the evolution of the system's many body density matrix is driven by the Schrodinger equation:
\begin{eqnarray}
  \partial_t  \rho(t) =-{i\over \hbar}[{\cal H}(t),\rho(t)].
\end{eqnarray}
where ${\cal H}(t)$ is the many body Hamiltonian.
For the two point correlation function, defined as:
\begin{eqnarray}
G_{{\bf r}{\bf r'}}(t)=\langle a_{\bf r}^\dag a_{\bf r'} \rangle=Tr (\rho(t)a_{\bf r}^\dag a_{\bf r'})
\end{eqnarray}
it follows that
\begin{eqnarray}
  \partial_t  G_{{\bf r}{\bf r'}}(t) =-{i\over \hbar} Tr ([{\cal H}(t),\rho(t)]a_{\bf r}^\dag a_{\bf r'})=-{i\over \hbar}Tr \rho(t)[a_{\bf r}^\dag a_{\bf r'},{\cal H}(t)]).
\end{eqnarray}
For a general interacting Hamiltonian, the equation above is very complicated, and $\partial_t  G_{{\bf r}{\bf r'}}(t)$ involves high order correlations through the right hand side of the equation, and we do not have a closed equation for the matrix $G$. 

In the case where particles are non-interacting Fermions (for example neutral fermionic atoms, or when  electron-electron interactions are screened), a closed equation for $G$ is available. In such a situation, the  Hamiltonian is quadratic in creation/annihilation operators, i.e. of the form ${\cal H}(t)=\sum_{{\bf x}{\bf y}} H_{{\bf x}{\bf y}}(t) a_{\bf x}^\dag a_{\bf y}$. Using the cannonical anti-commutation relations $\{a_{\bf x}^\dag, a_{\bf y}\}=\delta_{\bf x y}$ and $\{a_{\bf x}, a_{\bf y}\}=0$, the commutator in the equation of motion for $G$ is of the form
\begin{eqnarray}
    [{\cal H}(t),a_{\bf r}^\dag a_{\bf r'}]=\sum_{{\bf x}{\bf y}} H_{{\bf x}{\bf y}}(t) [a_{\bf x}^\dag a_{\bf y},a_{\bf r}^\dag a_{\bf r'}]=\sum_{{\bf x}{\bf y}} H_{{\bf x}{\bf y}}(t)(\delta_{\bf y r}a_{\bf x}^\dag a_{\bf r'}-\delta_{\bf x r'}  a_{\bf r}^\dag a_{\bf y}) 
\end{eqnarray} 
leading to:
\begin{eqnarray}& \label{eq:dtGdetails}
   \partial_t  G_{{\bf r}{\bf r'}}(t) ={i\over \hbar}Tr \rho(t)[{\cal H}(t),a_{\bf r}^\dag a_{\bf r'}])={i\over \hbar}\sum_{{\bf x}} H_{{\bf x}{\bf r}}(t) Tr \rho(t)(a_{\bf x}^\dag a_{\bf r'})-{i\over \hbar}\sum_{{\bf y}} H_{{\bf r'}{\bf y}}(t) Tr \rho(t)( a_{\bf r}^\dag a_{\bf y}))\\ & = {i\over \hbar}\sum_{{\bf x}} H_{{\bf x}{\bf r}}(t) G_{\bf x r'}(t)-{i\over \hbar}\sum_{{\bf y}} H_{{\bf r'}{\bf y}}(t) G_{\bf r  y}(t)={i\over \hbar} ([H^T(t),G(t)])_{{\bf r}{\bf r'}}
\end{eqnarray} 
yielding a closed equation for $G$. In particular, the local density at point ${\bf r}$ which is $G_{\bf r r}(t)$ evolves as: 
\begin{eqnarray}& 
   \partial_t  G_{{\bf r}{\bf r}}(t) ={i\over \hbar} ([H^T(t),G(t)])_{{\bf r}{\bf r}}.
\end{eqnarray} 
It is important to note, that this equation is not a closed equation for the density, since it involves off diagonal terms in $G$, thus to find the evolution of the density, the full Eq \eqref{eq:dtGdetails} must be solved. 
For a {\it real} Hamiltonian, as discussed in this paper, this equation is solved, for any time $\tau$ by 
\begin{eqnarray}\label{KUpotential}
    G(t+\tau)={\cal T}e^{{i\over \hbar}\int_{t}^{t+\tau}H(s)ds} G ({\cal T}e^{{i\over \hbar}\int_{t}^{t+\tau}H(s)ds} )^{\dag}=K_U(G)
\end{eqnarray}
as described in Eq. \eqref{K_Free}.

The discrete-time co-moving steady state equation on the lattice  $G=S^{\dag}{\cal K}(G)S$, Eq. \eqref{steadystateGeneral}, states that as the potential (or another type of disturbance) moves to the next lattice site, $G$ remains invariant, up to a shift of the coordinates co-moving with the disturbance.
Written explicitly in coordinate representation, Eq. \eqref{steadystateGeneral} reads $G_{\bf r r'}=[S^{\dag}{\cal K}(G)S]_{\bf r r'},$ or, equivalently, using that for momentum states, $S\ket{\bf k}=e^{-i a \mathbf{w} \cdot \mathbf{k} }\ket{\bf k}$, we have
\begin{eqnarray}\label{eq:steadyMomentum}
    G_{\bf k k'}=[S^{\dag}{\cal K}(G)S]_{\bf k k'}=	e^{i a \mathbf{w} \cdot (\mathbf{k} - \mathbf{k'})} [{\cal K}(G)]_{\mathbf{k}\mathbf{k'}}
\end{eqnarray}
\old
Let us write $G=G_0 +\delta G$, where $G_0$ is the initial steady state before turning on the traveling perturbation.
Using the form \eqref{KUpotential}, taking the Hamiltonian $H$ between steps to be of the form $H_0+V$, the steady state equation \eqref{eq:steadyMomentum} is explicitly given by:
	\begin{eqnarray}\label{eq:steadykapp}
	e^{i a \mathbf{w} \cdot (\mathbf{k} - \mathbf{k'})} \mel{\mathbf{k}}{e^{i \tau (H_0 + V)} (G_0 + \delta G) e^{-i \tau (H_0 + V)}}{\mathbf{k'}} = \mel{\mathbf{k}}{G_0 + \delta G}{\mathbf{k'}}
	\end{eqnarray}
	Note that at this no approximation has been made up to this point. 
Since we are only perturbing the free evolution by a local potential, we can assume the steady state $G$ will be close to the steady state of free evolution, $G_0$, and thus $\delta G$ is assumed a small perturbation. To proceed,  we now use perturbation theory in Eq. \eqref{eq:steadykapp}, by expanding to lowest order in $V$. Namely, we use the expansion :
\begin{eqnarray}
e^{-i \tau (H_0+V)} \approx e^{-i \tau H_0} + ie^{-i \tau H_0} \int_{0}^{\tau} ds e^{i s H_0} V e^{-i s H_0}
\end{eqnarray}
and keep terms up to linear order in $V$ and in $\delta G$. The resulting equation is:
	\begin{eqnarray}
 e^{i a \mathbf{w} \cdot (\mathbf{k} - \mathbf{k'})} \{ \mel{\mathbf{k}}{G_0}{\mathbf{k'}} + \mel{\mathbf{k}}{e^{i \tau H_0} \delta G e^{-i \tau H_0}}{\mathbf{k'}} + \mel{\mathbf{k}}{e^{i \tau H_0} \left( i \int_{0}^{\tau} ds e^{i s H_0} \comm{G_0}{V} e^{-i s H_0} \right) e^{-i \tau H_0}}{\mathbf{k'}} \}\\	= \mel{\mathbf{k}}{G_0}{\mathbf{k'}} + \mel{\mathbf{k}}{\delta G}{\mathbf{k'}}.
	\end{eqnarray}
Next we note that without the perturbation, $G_0$ is assumed to be a steady state of free evolution. Since $H$ is transnational invariant,  $G_0$ can be taken to be diagonal in momentum space, therefore we can set $\left( e^{i a \mathbf{w} \cdot (\mathbf{k} - \mathbf{k'})} - 1\right) \mel{\mathbf{k}}{G_0}{\mathbf{k'}}=0$.
Doing the $s$ integral, 
	\[
	    i \int_0^\tau ds e^{i s \left[ \varepsilon(\mathbf{k}) - \varepsilon(\mathbf{k'}) \right]} = \frac{ e^{i \tau \left[ \varepsilon(\mathbf{k}) - \varepsilon(\mathbf{k'}) \right]} - 1 }{\varepsilon(\mathbf{k}) - \varepsilon(\mathbf{k'})}
	\]
	we find  \old
	\begin{eqnarray}
 \left(e^{i a \mathbf{w} \cdot (\mathbf{k} - \mathbf{k'})} e^{i \tau \left[ \varepsilon(\mathbf{k}) - \varepsilon(\mathbf{k'}) \right]}  - 1 \right) \mel{\mathbf{k}}{\delta G}{\mathbf{k'}}
		=  \frac{\left(e^{i \tau \left[ \varepsilon(\mathbf{k}) - \varepsilon(\mathbf{k'}) \right]} - 1\right) e^{i a \mathbf{w} \cdot (\mathbf{k} - \mathbf{k'})} e^{i \tau \left[ \varepsilon(\mathbf{k}) - \varepsilon(\mathbf{k'}) \right]}}{\varepsilon(\mathbf{k}) - \varepsilon(\mathbf{k'})} \mel{\mathbf{k}}{\comm{G_0}{V}}{\mathbf{k'}}
	\end{eqnarray}
At zero temperature
	\begin{eqnarray} &
\mel{\mathbf{k}}{G_0 V}{\mathbf{k'}} = G_0(\mathbf{k})\mel{\mathbf{k}}{V}{\mathbf{k'}}= \Theta(\varepsilon_f - \varepsilon(\mathbf{k})) \sum_{\mathbf{r}} \braket{\mathbf{k}}{\mathbf{r}} V(\mathbf{r})  \braket{\mathbf{r}}{\mathbf{k'}}=\\ &  \frac{V}{ vol} \Theta(\varepsilon_f - \varepsilon(\mathbf{k})) e^{i \mathbf{r_0} \cdot (\mathbf{k} - \mathbf{k'})} ,
	\end{eqnarray}
where $vol$ is the system volume. Thus,
	\begin{eqnarray}\label{eqdeltak}
		\mel{\mathbf{k}}{\delta G}{\mathbf{k'}} = 	 \frac{V \left(e^{i \tau \left[ \varepsilon(\mathbf{k}) - \varepsilon(\mathbf{k'}) \right]} - 1\right) e^{i a \mathbf{w} \cdot (\mathbf{k} - \mathbf{k'})} e^{i \tau \left[ \varepsilon(\mathbf{k}) - \varepsilon(\mathbf{k'}) \right]}}{vol
			[\varepsilon(\mathbf{k}) - \varepsilon(\mathbf{k'})]\left(e^{i a \mathbf{w} \cdot (\mathbf{k} - \mathbf{k'})} e^{i \tau \left[ \varepsilon(\mathbf{k}) - \varepsilon(\mathbf{k'}) \right]}  - 1 \right) } e^{i \mathbf{r_0} \cdot (\mathbf{k} - \mathbf{k'})} \left[ \Theta(\varepsilon_f - \varepsilon(\mathbf{k})) - \Theta(\varepsilon_f - \varepsilon(\mathbf{k'})) \right]
	\end{eqnarray}
Finally, density variation in the co-moving non-equilibrium  steady state is given by Fourier transforming \eqref{eqdeltak} back to real space and taking the diagonal element, \old
	\begin{gather}
		\mel{\mathbf{r}}{\delta G}{\mathbf{r}} = \nonumber\\
		\frac{V \tau a^4}{(2 \pi)^4}\iint_{-\pi/a}^{\pi/a} d\mathbf{k} d\mathbf{k'}
	 \left[ \frac{e^{i \tau \left[ \varepsilon(\mathbf{k}) - \varepsilon(\mathbf{k'}) \right]} - 1}{\tau(\varepsilon(\mathbf{k}) - \varepsilon(\mathbf{k'}))} \right]
	 \left[\frac{1}{1 - e^{-i a \mathbf{w} \cdot (\mathbf{k} - \mathbf{k'})} e^{-i \tau \left[ \varepsilon(\mathbf{k}) - \varepsilon(\mathbf{k'}) \right]}} \right]
		e^{i (\mathbf{r_0}-\mathbf{r}) \cdot (\mathbf{k} - \mathbf{k'})}
	\left[ \Theta(\varepsilon_f - \varepsilon(\mathbf{k})) - \Theta(\varepsilon_f - \varepsilon(\mathbf{k'})) \right]\label{rspaceden}
	\end{gather}

\begin{figure}
    \centering
    \includegraphics[width=\linewidth]{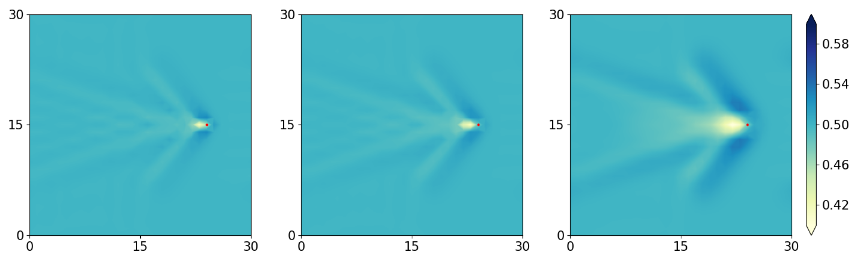}
    \caption{A Gaussian potential at half-filling and $\alpha = 1.7$.  From left to right, the standard deviation of the Gaussian in terms of lattice spacing, a, is point potential, $0.5 a$, and $a$.}
    \label{fig: Gaussian}
\end{figure}

\begin{figure}
    \centering
    \includegraphics[width=0.75\linewidth]{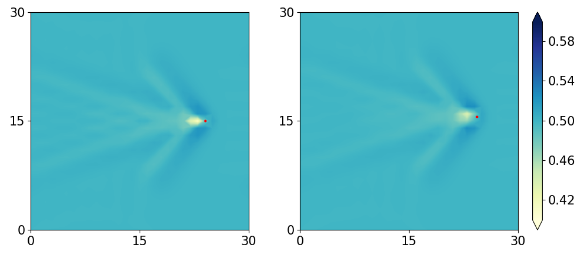}
    \caption{A Gaussian potential at half-filling, $\alpha = 1.7$, and $0.5 a$ standard of deviation where $a$ is the lattice spacing.  Left is a Gaussian starting directly on lattice site (8,15).  Right is a Gaussian starting in-between lattice sites at (8.3,15.6).}
    \label{fig: Shifted Gaussian}
\end{figure}

In experiments the point potential may be realized by a broader beam. As we are looking for far field wake patterns, we do not expect local structure of the disturbance to have a significant effect.  For example, if the point potential was instead Gaussian, the term $e^{i (\mathbf{r_0}-\mathbf{r}) \cdot (\mathbf{k} - \mathbf{k'})}$ in Eq.~\eqref{rspaceden} would be replaced by $e^{- \frac{\sigma^2}{2} |\mathbf{k} - \mathbf{k'}|^2 + i (\mathbf{r_0}-\mathbf{r}) \cdot (\mathbf{k} - \mathbf{k'})}$ where $\sigma^2$ is the variance of the Gaussian potential.  Note, however, that when calculating the wake geometry we look far away from the potential source ($|\mathbf{r_0}-\mathbf{r}|$ large) and also need only to consider terms where $|\mathbf{k} - \mathbf{k'}|$ is small since these $\mathbf{k}$,$\mathbf{k'}$ dominate the integral.  Hence, we can safely neglect the $e^{- \frac{\sigma^2}{2} |\mathbf{k} - \mathbf{k'}|^2}$ term so long as $|\mathbf{r_0}-\mathbf{r}|>>\sigma$ and thus find that the wake geometry of the Gaussian potential is equivalent to that of the point potential.  In Figs. \ref{fig: Gaussian} and \ref{fig: Shifted Gaussian} we simulate potential wakes for a selection of Gaussian potentials and find that the wake geometry is indeed equivalent to that of the point potential.
\\
(2) {\bf Moving detection/extraction.}
The co-moving steady state equation reads $G=S^\dagger \cal{K}(G)$. Written explicitly in momentum space, using \eqref{extraction detection}, we have:
    \begin{gather}\label{eq dist_ext}
        e^{i a \mathbf{w} \cdot (\mathbf{k} - \mathbf{k'})} e^{i \tau \left[ \varepsilon(\mathbf{k}) - \varepsilon(\mathbf{k'}) \right]} \{ \mel{k}{G}{k'} - \gamma \mel{\mathbf{k}}{ \{ G,P \}}{\mathbf{k'}} + \xi \mel{\mathbf{k}}{P G P}{\mathbf{k'}} \} = \mel{\mathbf{k}}{G}{\mathbf{k'}}
    \end{gather}
Assuming that $\gamma \ll 1$, $G \approx G_0 + \delta G$, and zero temperature, 
    \begin{align}
        \gamma \mel{\mathbf{k}}{ G P }{\mathbf{k'}} &
        \approx\gamma \mel{\mathbf{k}}{G_0 P}{\mathbf{k'}} = \frac{\gamma}{vol} \Theta(\varepsilon_f - \varepsilon(\mathbf{k})) e^{i \mathbf{r_0} \cdot (\mathbf{k} - \mathbf{k'})}
    \end{align}
    Hence,
    \begin{gather}
        \label{eq: GP + PG}
        \gamma \mel{\mathbf{k}}{ \{ G,P \}}{\mathbf{k'}} \approx \frac{\gamma}{vol} e^{i \mathbf{r_0} \cdot (\mathbf{k} - \mathbf{k'})} \left[ \Theta(\varepsilon_f - \varepsilon(\mathbf{k})) + \Theta(\varepsilon_f - \varepsilon(\mathbf{k'})) \right]~.
    \end{gather}
 Now turning to the $PGP$ term,
    \begin{gather}
        \label{eq: PGP}
        \mel{\mathbf{k}}{P G_0 P}{\mathbf{k'}} = \frac{a^2 vol }{(2 \pi)^2} \int d\mathbf{q} \Theta(\varepsilon_f - \varepsilon(\mathbf{q})) \mel{\mathbf{k}}{P}{\mathbf{q}} \mel{\mathbf{q}}{P}{\mathbf{k'}}
        = \frac{a^2 vol }{(2 \pi)^2} e^{i \mathbf{r_0} \cdot (\mathbf{k} - \mathbf{k'})} \int d\mathbf{q} \Theta(\varepsilon_f - \varepsilon(\mathbf{q})) = \frac{\rho_f}{vol} e^{i \mathbf{r_0} \cdot (\mathbf{k} - \mathbf{k'})} ~.
    \end{gather}
    where $\rho_f$ is the density of fermions for $G_0$.  

 Plugging Eqs. \eqref{eq: GP + PG} and \eqref{eq: PGP} into Eq. \eqref{eq dist_ext}, we get
    \begin{gather}
        \mel{\mathbf{k}}{\delta G}{\mathbf{k'}} = \frac{\gamma}{vol} \left( \frac{e^{i a \mathbf{w} \cdot (\mathbf{k} - \mathbf{k'})} e^{i \tau \left[ \varepsilon(\mathbf{k}) - \varepsilon(\mathbf{k'}) \right]}}{e^{i a \mathbf{w} \cdot (\mathbf{k} - \mathbf{k'})} e^{i \tau  \left[ \varepsilon(\mathbf{k}) - \varepsilon(\mathbf{k'}) \right]} - 1} \right) \left[ \frac{\xi \rho_f}{\gamma} - \Theta(\varepsilon_f - \varepsilon(\mathbf{k})) - \Theta(\varepsilon_f - \varepsilon(\mathbf{k'})) \right] e^{i \mathbf{r_0} \cdot (\mathbf{k} - \mathbf{k'})}~.
    \end{gather}
Finally, the local density variation is given by:
    \begin{gather}
    \mel{\mathbf{r}}{\delta G}{\mathbf{r}} = \frac{\gamma a^4}{(2\pi)^4}  \iint_{-\pi/a}^{\pi/a} d\mathbf{k} d\mathbf{k'}   \left( \frac{1}{1 - e^{-i a \mathbf{w} \cdot (\mathbf{k} - \mathbf{k'})} e^{-i \tau \left[ \varepsilon(\mathbf{k}) - \varepsilon(\mathbf{k'}) \right]}} \right) e^{i (\mathbf{r_0} - \mathbf{r}) \cdot (\mathbf{k} - \mathbf{k'})}  \left[ \frac{\xi \rho_f}{\gamma} - \Theta(\varepsilon_f - \varepsilon(\mathbf{k})) - \Theta(\varepsilon_f - \varepsilon(\mathbf{k'})) \right] 
    \end{gather}
which is Eq.  \eqref{eq: detection final} in the main text.

\section{Non-perturbative Results}
In this section, we show that no detection wake is created at $\rho_f = \frac{1}{2}$ and that the difference between detection and extraction is temperature independent  even non-perturbatively.

We start by looking at a series of non-perturbative detections on $G_0$.  A single detection at site r and evolution for time $\tau$ is 

\begin{eqnarray}
    & G = U P_r G_0 P_r U^\dagger + U P_r^\perp G_0 P_r^\perp U^\dagger \\ \nonumber & \equiv \sum_{a = \{0,1\}} P_r^a (\tau) G_0 P_r^a (\tau)
\end{eqnarray}

since $\comm{U}{G_0} = 0$ and where $P^0 \equiv P$, $P^1 \equiv P^\perp$, and $U P U^\dagger \equiv P (\tau)$.  

Hence, after doing a series of $m$ measurements, we have

\begin{eqnarray}
    G = \sum_{a_1,a_2,...,a_m} \left\{ \left[ \prod_{n=m,m-1,...,1} P_{r_n}^{a_n} ((m-n + 1)\tau)\right] G_0 \left[ \prod_{n=1,2,...,m} P_{r_n}^{a_n} ((m-n + 1)\tau)\right] \right\}
    \label{Eq: Non-pert G Detection unsimplified}
\end{eqnarray}

Looking at the diagonal of $G$ in real space and inserting a resolution of identity, $\int d\mathbf{q_n} \dyad{\mathbf{q_n}}{\mathbf{q_n}}$, to the right of every $P_{r_n}^{a_n}$ sitting in the first term in brackets in Eq. \eqref{Eq: Non-pert G Detection unsimplified} and inserting $\int d\mathbf{q'_n} \dyad{\mathbf{q'_n}}{\mathbf{q'_n}}$ to the left of every $P_{r_n}^{a_n}$ sitting in the second bracketed term in Eq. \eqref{Eq: Non-pert G Detection unsimplified} we find

\begin{align}
    \mel{\mathbf{r}}{G}{\mathbf{r}} \equiv \zeta_m (\mu) = \int d\mathbf{k} d\mathbf{k'} e^{-i \mathbf{r} \cdot (\mathbf{k} - \mathbf{k'})} & \int dQ dQ' \sum_{a_1,a_2,...,a_m} \left\{ \mel{\mathbf{k}}{P_{r_m}^{a_m} (\tau)}{\mathbf{q_m}} \mel{\mathbf{q_m}}{P_{r_{m-1}}^{a_{m-1}} (2 \tau)}{\mathbf{q_{m-1}}} ...  \right. \\ \nonumber \times & \left. \mel{\mathbf{q_2}}{P_{r_{1}}^{a_{1}} (m \tau)}{\mathbf{q_1}} \mel{\mathbf{q_1}}{G_0}{\mathbf{q'_1}} \mel{\mathbf{q'_1}}{P_{r_{1}}^{a_{1}} (m \tau)}{\mathbf{q'_2}}... \mel{\mathbf{q'_m}}{P_{r_m}^{a_m} (\tau)}{\mathbf{k'}} \right\}
\end{align}

where $\mu$ is the chemical potential.  Now, focusing on only terms directly dependent on $\mathbf{q_1}$,$\mathbf{q'_1}$ and denoting all other terms by $B$, we find 

\begin{eqnarray}
    & \zeta_m (\mu) = B \int d\mathbf{q_1} d\mathbf{q'_1} \sum_{a_1} \mel{\mathbf{q_2}}{P_{r_{1}}^{a_{1}} (m \tau)}{\mathbf{q_1}} \mel{\mathbf{q_1}}{G_0}{\mathbf{q'_1}} \mel{\mathbf{q'_1}}{P_{r_{1}}^{a_{1}} (m \tau)}{\mathbf{q'_2}} \\ \nonumber & = B \int d\mathbf{q_1} \sum_{a_1} \mel{\mathbf{q_2}}{P_{r_{1}}^{a_{1}} (m \tau)}{\mathbf{q_1}}  F_\mu(\varepsilon(\mathbf{q_1}))
    \mel{\mathbf{q_1}}{P_{r_{1}}^{a_{1}} (m \tau)}{\mathbf{q'_2}} \\ \nonumber & = B \int d\mathbf{q_1} e^{i m \tau \left[ \varepsilon(\mathbf{q_2}) - \varepsilon(\mathbf{q'_2}) \right]}  F_\mu(\varepsilon(\mathbf{q_1})) \left[ \delta_{\mathbf{q_1} \mathbf{q_2}} \delta_{\mathbf{q_1} \mathbf{q'_2}}  + e^{i \mathbf{r_1} \cdot (\mathbf{q_2} - \mathbf{q'_2})} \left( 2 - \delta_{\mathbf{q_1} \mathbf{q_2}} - \delta_{\mathbf{q_1} \mathbf{q'_2}} \right) \right] \\ \nonumber & = \zeta_{m-1} (\mu) + B  \int d\mathbf{q_1} e^{i m \tau \left[ \varepsilon(\mathbf{q_2}) - \varepsilon(\mathbf{q'_2}) \right]}  F_\mu(\varepsilon(\mathbf{q_1})) e^{i \mathbf{r_1} \cdot (\mathbf{q_2} - \mathbf{q'_2})} \left( 2 - \delta_{\mathbf{q_1} \mathbf{q_2}} - \delta_{\mathbf{q_1} \mathbf{q'_2}} \right) \\ \nonumber & = \zeta_{m-1} (\mu) + B e^{i m \tau \left[ \varepsilon(\mathbf{q_2}) - \varepsilon(\mathbf{q'_2}) \right]}  e^{i \mathbf{r_1} \cdot (\mathbf{q_2} - \mathbf{q'_2})} \left( 2 \rho_f - F_\mu(\varepsilon(\mathbf{q_2})) - F_\mu(\varepsilon(\mathbf{q'_2})) \right)
    \label{Eq: non-pert simplification}
\end{eqnarray}

Now, we look at $\zeta_m(\mu) + \zeta_m(-\mu)$ in a way analogous to Eq. \eqref{Eq: mu plus minus mu} in the main text.  Here, we find

\begin{eqnarray}
    & \zeta_m(\mu) + \zeta_m(-\mu) = \zeta_{m-1}(\mu) + \zeta_{m-1}(-\mu)  \label{Eq. non-pert mu plus minus mu} \\ \nonumber & + B \left\{ e^{i m \tau \left[ \varepsilon(\mathbf{q_2}) - \varepsilon(\mathbf{q'_2}) \right]}  e^{i \mathbf{r_1} \cdot (\mathbf{q_2} - \mathbf{q'_2})} \left( 2 - F_\mu(\varepsilon(\mathbf{q_2}))  - F_{\mu}(\varepsilon(\mathbf{q'_2})) - F_{-\mu}(\varepsilon(\mathbf{q_2})) - F_{-\mu}(\varepsilon(\mathbf{q'_2})) \right) \right\}
\end{eqnarray}

Note, the real part of the term in braces in Eq. \eqref{Eq. non-pert mu plus minus mu} is anti-symmetric under the transformation $\mathbf{q} \rightarrow {\cal M} (\mathbf{q})$, where $\mathbf{q}$ here represents all $\mathbf{q_n}$, $\mathbf{q'_n}$, $\mathbf{k}$, and $\mathbf{k'}$.  Let us now look at the term $B$.  Note, explicitly, 

\begin{equation}
    B = \int dQ_1 dQ'_1  e^{-i \mathbf{r} \cdot (\mathbf{q_{m+1}} - \mathbf{q'_{m+1}})}  \sum_{a_2,...,a_m} \prod_{n=m+1,m,...,2} \mel{\mathbf{q_n}}{P_{r_{n-1}}^{a_{n-1}} ((m-n + 2)\tau)}{\mathbf{q_{n-1}}} \mel{\mathbf{q'_{n-1}}}{P_{r_{n-1}}^{a_n} ((m-n + 2)\tau)}{\mathbf{q'_{n}}}
\end{equation}

where $dQ_1$ and $dQ'_1$ are defined by $\prod_{n=2,...,m+1} d\mathbf{q_n}$ and $\prod_{n=2,...,m+1} d\mathbf{q'_n}$ respectively.  Also, here we define $\mathbf{q_{m+1}} \equiv \mathbf{k}$ and $\mathbf{q'_{m+1}} \equiv \mathbf{k'}$. 

Simplifying $B$, we find

\begin{eqnarray}
    & B = \int dQ_1 dQ'_1 e^{-i \mathbf{r} \cdot (\mathbf{q_{m+1}} - \mathbf{q'_{m+1}})} \prod_{n=m+1,m,...,2} e^{i (m-n+2) \tau \left[ \varepsilon(\mathbf{q_n}) - \varepsilon(\mathbf{q_{n-1}}) + \varepsilon(\mathbf{q'_{n-1}}) - \varepsilon(\mathbf{q'_{n}}) \right]} \label{Eq. non-pert B final} \\ \nonumber & \times \left[ \delta_{\mathbf{q_n} \mathbf{q_{n-1}}} \delta_{\mathbf{q'_n} \mathbf{q'_{n-1}}} - e^{i \mathbf{r_{n-1}} \cdot (\mathbf{q_n} - \mathbf{q_{n-1}})} \delta_{\mathbf{q'_n} \mathbf{q'_{n-1}}} - e^{i \mathbf{r_{n-1}} \cdot (\mathbf{q'_{n-1}} - \mathbf{q'_{n}})} \delta_{\mathbf{q_n} \mathbf{q_{n-1}}} + 2 e^{i \mathbf{r_{n-1}} \cdot (\mathbf{q_n} - \mathbf{q_{n-1}})} e^{i \mathbf{r_{n-1}} \cdot (\mathbf{q'_{n-1}} - \mathbf{q'_{n}})} \right]
\end{eqnarray}

It can now be seen from Eq. \eqref{Eq. non-pert B final} that B is symmetric under the transformation $\mathbf{q} \rightarrow {\cal M} (\mathbf{q})$.  Since $B$ is symmetric and the term in braces in Eq. \eqref{Eq. non-pert mu plus minus mu} is anti-symmetric, we find

\begin{equation}
    \zeta_m(\mu) + \zeta_m(-\mu) = \zeta_{m-1}(\mu) + \zeta_{m-1}(-\mu)
\end{equation}

Thus, $\zeta_m(\mu) + \zeta_m(-\mu) = \zeta_{0}(\mu) + \zeta_{0}(-\mu) = 1$ and 

\begin{equation}
    \mel{\mathbf{r}}{G}{\mathbf{r}}_\mu = 1 - \mel{\mathbf{r}}{G}{\mathbf{r}}_{-\mu}
\end{equation}

i.e. the detection wake for a chemical potential of $\mu$ is one minus the detection wake for a chemical potential of $-\mu$.  Hence, when $\mu = 0$ there is no detection wake.  We emphasize that this result assumed no particular path for the moving detector.

Turning to a moving extractor, note that for the difference between the extractor and detector wake, we get Eq.~\eqref{Eq: Non-pert G Detection unsimplified} where we set $a_1,a_2,...,a_m = 0$.  Thus, Eq. \eqref{Eq: non-pert simplification} becomes

\begin{eqnarray}
    & \zeta_m = B \int d\mathbf{q_1} e^{i m \tau \left[ \varepsilon(\mathbf{q_2}) - \varepsilon(\mathbf{q'_2}) \right]} \delta_{\mathbf{q_1} \mathbf{q'_1}} F(\varepsilon(\mathbf{q_1})) e^{i \mathbf{r_1} \cdot (\mathbf{q_2} - \mathbf{q'_2})} = B e^{i m \tau \left[ \varepsilon(\mathbf{q_2}) - \varepsilon(\mathbf{q'_2}) \right]} e^{i \mathbf{r_1} \cdot (\mathbf{q_2} - \mathbf{q'_2})} \rho_f
\end{eqnarray}

Hence, the difference between a moving detector and moving extractor is temperature independent non-perturbatively.  Similar to the perturbative case, this implies that a moving particle extractor at $\rho_f = \frac{1}{2}$ is temperature independent.  Again, note that we have assumed no particular path for our moving particle extractor.

\begin{figure}
    \centering
    \includegraphics[width=0.5\linewidth]{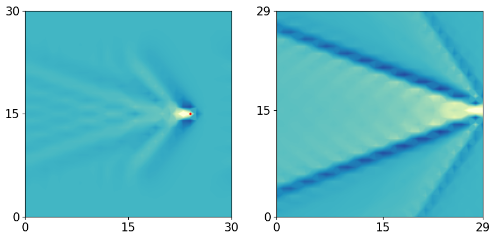}
    \caption{Comparison of potential wake at half-filling for $\alpha = 1.7$ for simulation (left) and numerical integration of Eq. \eqref{Eq. Pot Real Space All Quadrants} (right).}
    \label{fig:Int Approx}
\end{figure}

\end{document}